\documentclass[12pt]{article}

\usepackage{amssymb,amsmath, amsthm,amsopn,amsfonts,amsmath,mathrsfs,graphicx,cite,color}
\usepackage{verbatim} 

\newtheorem{theorem}{Theorem}

\newtheorem{remark}[theorem]{Remark}
\newtheorem{corollary}[theorem]{Corollary}

\newcommand{\RR}{\mathbb{R}}

\newcommand{\C}{\mathcal{C}}
\newcommand{\R}{\mathcal{R}}

\newcommand{\ma}{\mathbf{a}}

\newcommand{\malpha}{\boldsymbol{\alpha}}
\newcommand{\mbeta}{\boldsymbol{\beta}}
\newcommand{\momega}{\boldsymbol{\omega}}
\newcommand{\mx}{\mathbf{x}}
\newcommand{\my}{\mathbf{y}}
\newcommand{\mz}{\mathbf{z}}

\newcommand{\D}{\mathcal{D}}

\begin{document}
\title{Detecting small low emission radiating sources}
\author{Moritz Allmaras $^1$, David P. Darrow $^3$, Yulia Hristova $^2$,\\ Guido Kanschat $^1$, and Peter Kuchment $^1$
\\$^1$ Mathematics Department, Texas A\&M University, College Station, TX
\\$^2$ Institute for Mathematics and its Applications, Minneapolis, MN
\\$^3$ University of Texas Medical Branch, Galveston, TX}
\date{}
\maketitle
\abstract{
The article addresses the important for homeland security possibility of robust detection of geometrically small, low emission sources on a significantly stronger background. A technique of detecting such sources using Compton type cameras is developed and studied analytically and numerically, which is shown to have high sensitivity and specificity and also allows to assign confidence probabilities of the detection. The $2D$ case is considered in detail.}

\section{Introduction}

One of the missions of the Department of Homeland Security is to prevent smuggling of weapon-grade
nuclear materials (e.g., \cite{shield}). It is expected that such materials, unlike those needed for a ''dirty bomb'',
will have low emission rates and will be well shielded, so that very few gamma photons or neutrons would escape, and even less would be detected non-scattered (ballistic). An additional hurdle for the detection of illicit
nuclear substances is the strong natural radiation background. If the particles radiated by the source were physically distinct (e.g., in terms of their energies) from the ones prevalent in the background, then the detection would be easier. We thus assume that the particles from the source are identical to those coming from the background in terms of their nature, energies, etc., so the discrimination using such parameters is impossible. This leads to the situation when the signal to noise ratio (SNR), i.e. the count of the detected non-scattered particles emitted by the source versus the total number of detected particles, can be as low as $10^{-3}$, if not lower. It is rather clear that if only the locations of the hits are detected, with no information about the directions of the incoming particles available, there is no chance to detect the presence of such a weak source. Thus, one needs to employ detectors that can provide some directional information.
For instance, the $\gamma$-cameras most commonly used in SPECT (Single Photon Emission Computed Tomography) medical imaging \cite{Natterer_book,BGH}, are mechanically collimated, and thus they ``count'' the particles coming from a narrow cone of directions and discard the rest. However, mechanical collimation
dramatically reduces the particle count, and thus, in the case of an extremely low emission source, can essentially eliminate the useful signal (this difficulty, although in much less extreme form, arises also in SPECT). Typically, only one out of thousands emitted
photons is detected by a collimated camera. When dealing with low emission sources, one wants to
capture as many particles as possible, and thus mechanical collimation is unsuitable for the problem at hand.

Another option is to use the so called \textbf{Compton $\gamma$-cameras} (see Section \ref{S:compton}), which do not discard any incoming particles by collimation. The price one pays for this is that the directional information provided by such a camera is more limited. Namely, one obtains just a hollow (i.e., surface rather than solid) cone of possible directions rather than a single direction. Still, Compton cameras are good candidates for the applications we have in mind. While Compton cameras are available for detecting $\gamma$-photons, neutron detectors that provide a similar ``incoming cone'' information are being currently developed as well \cite{SpCharlton,shield}. The methods we will discuss do not depend upon a specific kind of particles; we will thus call all detectors that can provide the cone information \textbf{Compton type detectors}.

Even if some directional information is available, the task of detecting extremely low emission coming from a geometrically large source in the presence of a significant radiation background would be very hard, if not impossible. So, another important condition that we impose, besides availability of directional information, is the small size of the possible source, which is usually a safe assumption in the applications we are interested in.

Simulations of radiation from a small high-enriched uranium source placed in a cargo container suggest that only about $0.1\%$ of the signal received by detectors might be due to the ballistic (non-scattered) particles emitted from the source. The remaining $99.9\%$ of detected particles come either from extraneous (natural) sources, or from scattered source particles from the source \cite{Charlton}.

It is natural to try to use the rather standard SPECT techniques in the present situation, albeit the chances of success (and even the applicability of the tomographic models) are questionable. This circle of issues is discussed analytically and tested numerically in Section \ref{S:tomo}, where the conclusion is made that backprojection technique might be the best bet here, while the usual filtration parts in SPECT reconstructions do not do any good. This conclusion gets its foundation in the probabilistic discussions of Section \ref{S:prob} and then in numerical examples of Section \ref{S:backproj}. However, to check the viability of the approach, all these considerations dealt with the simplest, collimated detectors.
Section \ref{S:compton} is devoted to the case of Compton type cameras. Here tomographic and backprojection techniques are considered. The analysis of the $2D$ Compton camera case is provided in Section \ref{S:2DCom}. The corresponding numerical tests are conducted and discussed in Section \ref{S:ComEx}. The overall conclusion is that the suggested backprojection Compton type cameras techniques allow a robust detection of presence of geometrically small low emission sources with extremely low SNR, with confidence levels attached.

Like in the case of SPECT \cite{BGH}, one can try to use statistical methods. E.g., the recent paper \cite{Xun} describes a Bayesian approach to the same problem.

The paper ends with the sections devoted to remarks and conclusions, acknowledgments, and bibliography.

\section{Source detection using tomographic techniques}\label{S:tomo}

We will briefly present here some relevant mathematical formulas and conclusions from the SPECT version of emission tomography, which is the closest to the problem of our interest. One can find more details in \cite{Kuc_AMS,KLM,Natterer_new,Natterer_book,BGH}.

Let $f(x)$ be the intensity distribution of the sources of particles of certain type ($\gamma$-photons, neutrons, etc.) inside an object (e.g., cargo container, or a truck). Let also $\mu (x)$ be the  attenuation coefficient (usually not known the applications we have in mind). Then, in the approximation of sufficiently high emission rate, the radiation transport equation implies that the particle count per unit of time at a detector collimated in the direction $L$ is
\begin{equation}\label{E:Tmu}
T_\mu f(L)=\int_Lf(x)e^{-\int_{L_x}\mu (y)dy}dx.
\end{equation}
Here $L_x$ is the segment of the line $L$ between the emission
point $x$ and the detector and $dy$ denotes the standard linear
measure on $L$. The operator $T_\mu $ is said to be the {\bf
attenuated Radon (or X-ray) transform} (with attenuation $\mu (x)$) of the
function $f(x)$.
\begin{figure}[ht!]
\begin{center}
\scalebox{0.5}{\includegraphics{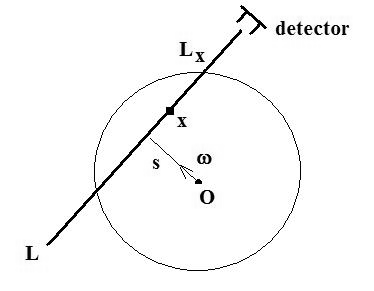}}
  \caption{The set-up of the SPECT emission tomography}\label{F:spect}
\end{center}
\end{figure}

We will fix an origin $O$ and assign the normal coordinates $(\omega,s)$ to any line $L$. Here $\omega = (\cos{\theta},\sin{\theta})$ is a unit vector orthogonal to $L$ and $s$ is the (signed) distance from the origin to $L$ (see Fig. \ref{F:spect}). The equation of this line is $x\cdot\omega=s$ and the line itself will be sometimes denoted as $L_{(\omega,s)}$.
The value of (\ref{E:Tmu}) provides the mathematical expectation of the counts at a detector per unit of time, which is not a good approximation of the measured data, if the emission rate is low. Let us still accept for the time being this integral-geometric formulation and see where it leads us.

The {\bf attenuated backprojection} operator with
attenuation $\nu(x,\omega)>0$ (which may or may not be related to $\mu(x)$) acting on data $g(\omega,s)$ is
\begin{equation}\label{E:Tmu_param}
  \left(T_{\nu}^{\#} g\right)(x)=\int_{|\omega|=1} g(\omega,x\cdot\omega)\nu(x,\omega)d\omega
\end{equation}
In particular, if $\nu=1$ (which we will assume in our experiments), this is the standard backprojection operator \cite{Kak,Natterer_new,Natterer_book}.

Let $H$ be the Hilbert transform acting in variable $s$ on functions $g(\omega,s)$:
$$
Hg(\omega,s)=\frac{1}{\pi}v.p.\int \frac{g(\omega,r)}{s-r}dr.
$$
Here $v.p.$ indicates that the integral is considered in the principal value meaning.

The following result holds:
\begin{theorem}(see details in \cite{KLM})
Let $f(x)$ be a function (our distribution of radiating sources) and $\mu$ be sufficiently smooth\footnote{The detection procedures
we will eventually come to, will not require smoothness of attenuation.},
then the operator
\begin{equation}\label{E:tomorec}
T^{\#}_{\nu}H\frac{d}{ds}T_{\mu}
\end{equation}
applied to $f$ preserves all singularities of $f$ and does not create new ones. In particular, if $f$ has a jump across an interface, the location of this jump is preserved in the image $T^{\#}_{\nu}H\frac{d}{ds}T_{\mu}f$.
\end{theorem}
Notice that $T^{\#}_{\nu}H\frac{d}{ds}T_{\mu}f$ is {\bf not} a reconstruction of $f$ from its measured projections $T_{\mu}f$ (which would be impossible anyway with the low signal strength and the attenuation $\mu$ being unknown), but it gives a correct reconstruction of all singularities (e.g., interfaces) of $f$.

Since we assume that our source is very small geometrically, it is rather singular, and thus one might expect that any guess for the unknown attenuation (e.g., that there is none) should reconstruct the location of the source correctly.

The operator
\begin{equation}\label{E:FBP}
T^{\#}_{1}H\frac{d}{ds}
\end{equation}
is exactly what is applied to the data in our experiments described below. We thus will not worry about the presence of an unknown attenuation.

The operator (\ref{E:FBP}) is often called a \textbf{filtered backprojection (FBP)}, where $H\frac{d}{ds}$ and $T^{\#}_{1}$ are the filtration and backprojection parts, correspondingly.

\begin{remark}
\begin{itemize}
\item
Since we are looking for a rather singular object, it is reasonable to recall that there is a way in tomography to emphasize singularities of the function to be reconstructed. This is what the so called {\bf local tomography} does (e.g., \cite{Faridani,KLM}). For instance, the singularities will look brighter if one replaces the filtration part $H\frac{d}{ds}$ in (\ref{E:FBP}) with a stronger filter $\frac{d^2}{ds^2}$:
\begin{equation}\label{E:loc}
T^{\#}_{1}\frac{d^2}{ds^2}.
\end{equation}
This option will also be explored.

\item In the detection problem we are discussing, different types of particles (e.g., $\gamma$-photons of different energies) will be present, and each of them could have different distribution $f_j$ of sources, as well as different attenuation distribution $\mu_j$ in the cargo. Thus, what we get if we lump all these particle counts together (which seems to be a strange thing to do), is the sum
$$
g=\sum_j T_{\mu_j}f_j.
$$
Can this complicate things? We claim that it should not.
Indeed, all these functions have the same boundary of the source as their interface.

One can derive now from the above theorem the following
\begin{corollary}Applying the same reconstruction
operator $T^{\#}_{\nu}H\frac{d}{ds}$ to the whole sum, one recovers the correct source interface.
\end{corollary}
\end{itemize}
\end{remark}

It is still not clear whether such a procedure will work in our problem, since there are at least two potentially serious obstacles:
\begin{itemize}
  \item The (attenuated) X-ray transform model does not apply when significant statistical noise is present, in particular for a very weak source.
  \item This model does not take into account presence of a very large (from standard tomographic point of view, enormous) background noise.
\end{itemize}

Still, we will try to apply this approach and see where it can lead us. We show below some sample results of numerical experimentation with the $2D$ and $3D$ X-ray transforms, as well as $3D$ Radon transform.

\subsection{$2D$ X-ray reconstructions}\label{SS:2DX}

Although the problem is three-dimensional, for the first trial we consider the $2D$ problem, which is more straightforward to handle numerically.

\begin{figure}[ht!]
\begin{center}
\scalebox{0.24}{\includegraphics{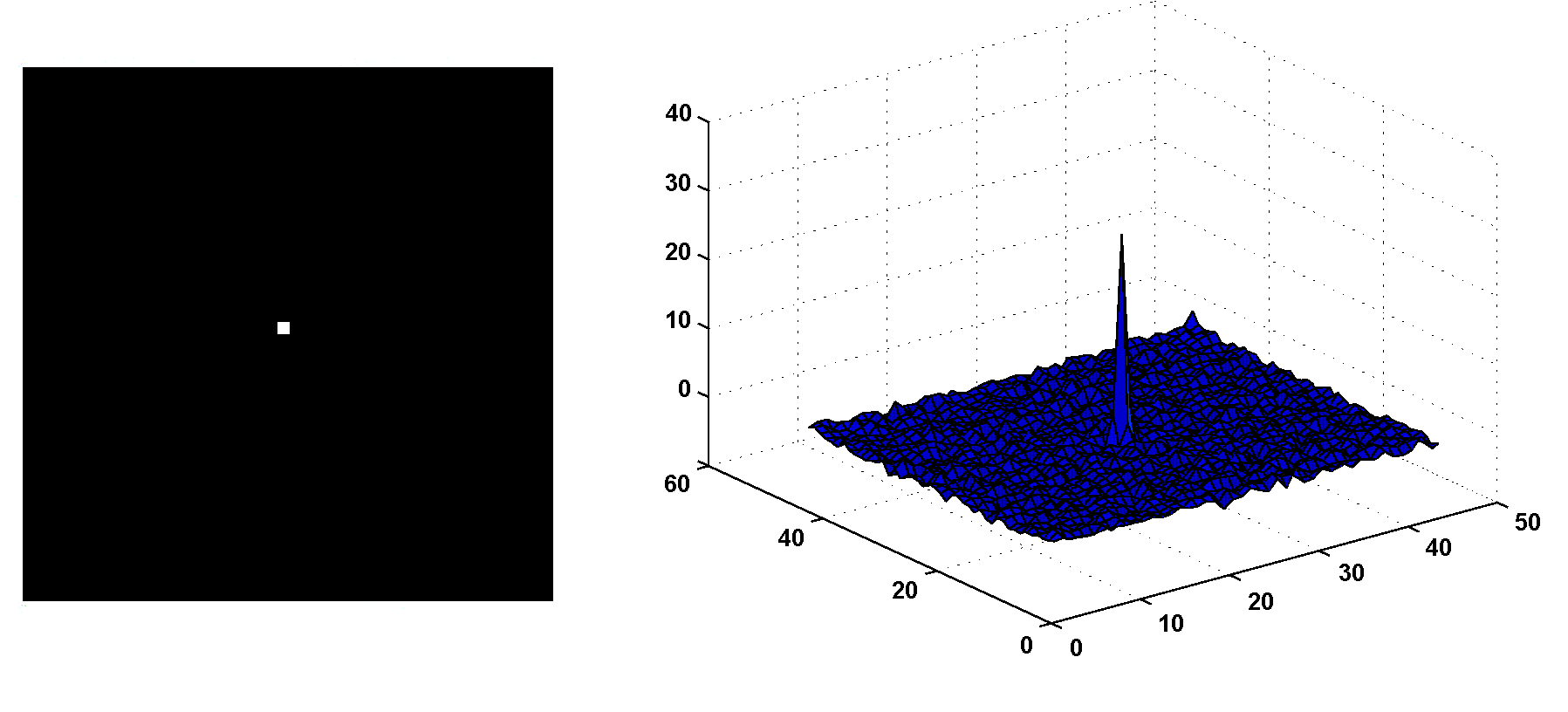}}
\caption{
Reconstruction with SNR $1\%$ (density and surface plots).}\label{F:30s}
\end{center}
\end{figure}

In the first reconstruction we have modeled random $\gamma$-photon emission from a shielded ball of radius $4$cm of $HEU235$ for realistic values of the emission and attenuation rates. The ball was placed inside in the center\footnote{The results for non-central locations of the source, as the further examples will show, are similar.} of a much (about $100$ times) larger ``cargo''. A uniform and isotropic random background was also added. The model included $64$ detectors, each of which could detect $64$ directional sectors (about $2.8$ degrees each). In this reconstruction we assumed the SNR to be $1\%$, i.e. $99\%$ of detected hits were generated by background, or scattered source particles.
Then, the described above tomographic $X$-ray transform inversion (see (\ref{E:tomorec}) with $\nu=1$) was applied to the resulting matrix of counts. The pictures below show the results of the reconstructions in $44\times44$ pixels (both the gray scale density and surface plots are presented). The result, shown in Fig.~\ref{F:30s}, shows clear detection of the source.

\begin{figure}[ht!]
\begin{center}
\scalebox{0.18}{\includegraphics{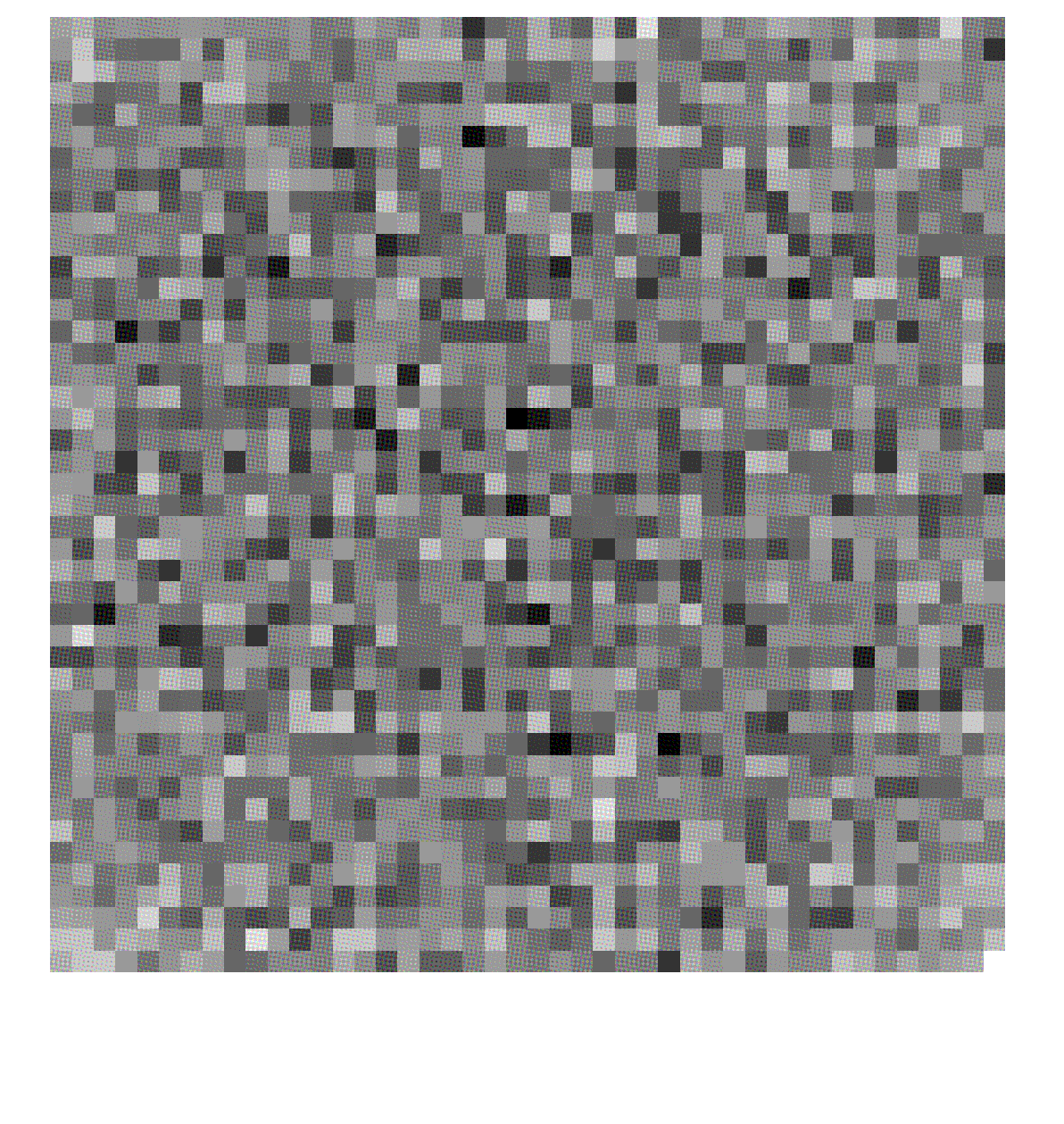}}
\scalebox{0.2}{\includegraphics{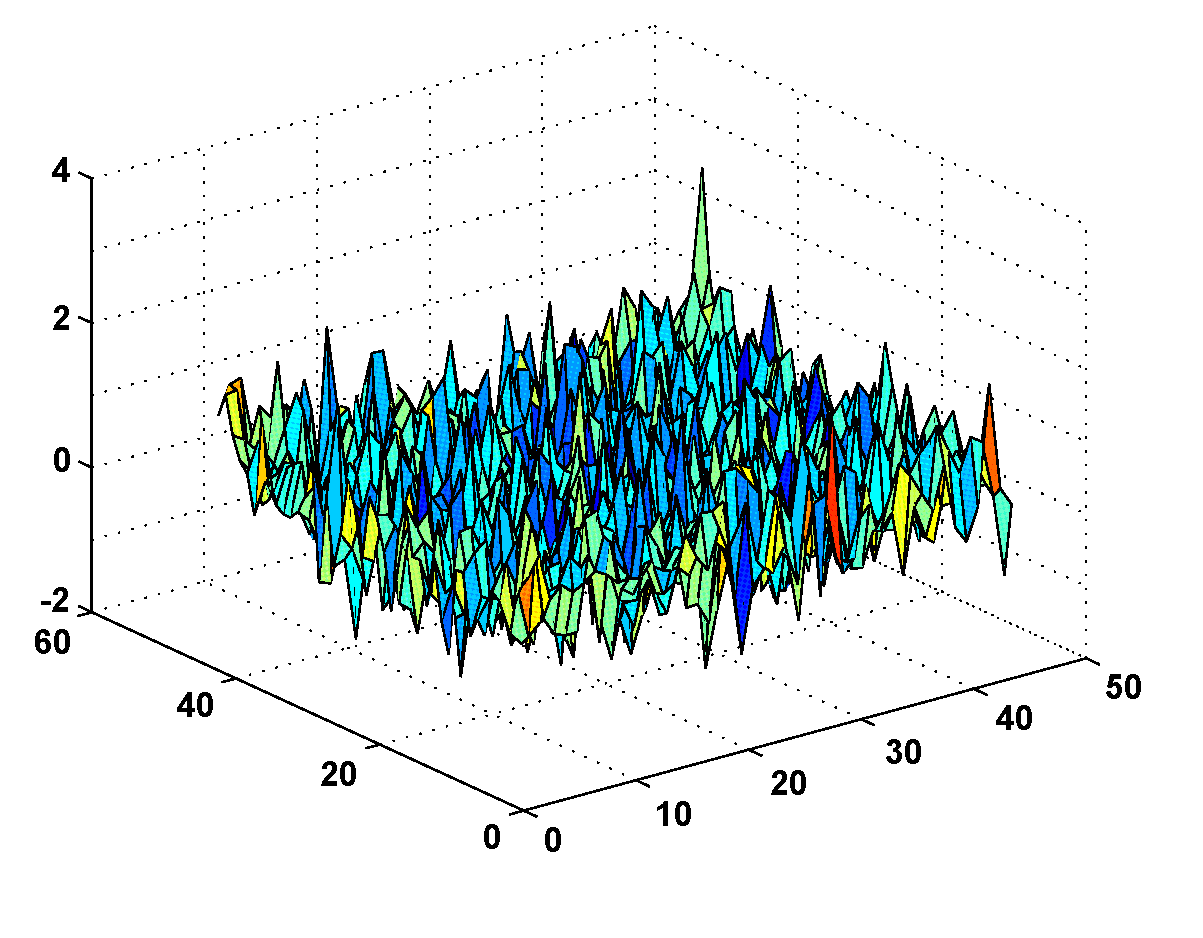}}\caption{
Unsuccessful FBP reconstruction with SNR $0.4\%$ and about $100$ ballistic source particles detected (density and surface plots).}\label{F:1s}
\end{center}
\end{figure}

We then try to lower the SNR towards our benchmark value $0.1\%$, and the success seems to evaporate. E.g., Fig. \ref{F:1s} shows the failed attempt of reconstruction with the  SNR being $0.4\%$ and with only about a hundred ballistic particles detected from the source.

Now let us consider a fixed signal-to-noise ratio ($0.1\%$) and varying total number of detected particles. We will first describe the set up for the numerical results presented in rest of this subsection.

\begin{figure}[ht!]
\begin{center}
\scalebox{0.4}{\includegraphics{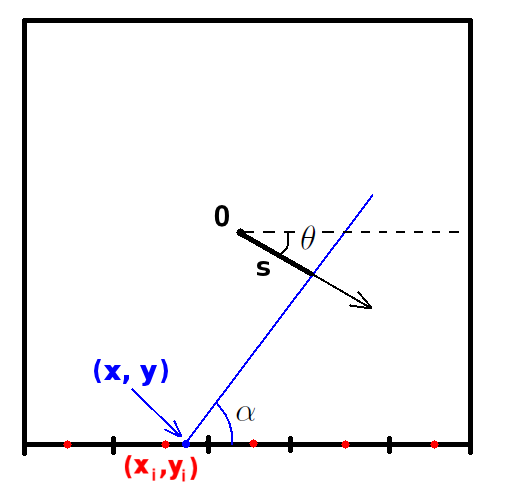}}
\caption{
A particle propagates along a random line with normal coordinates $(\theta,s)$ and collides with a detector array at the point $(x,y)$. We record the bin with center $(x_i,y_i)$ and the direction $\alpha = \theta+\pi/2$.}\label{F:setup}
\end{center}
\end{figure}

Four direction sensitive detector arrays, each consisting of $100$ detectors (bins), were placed along the sides of the square $[-1,1]^2$. Random background radiation was created by simulating particles propagating along straight lines in such a way that these lines were uniformly randomly distributed in the square. More precisely, to each particle we assigned a direction of propagation by choosing the normal coordinates $(\theta,s)$ of a line such that $\theta$ and $s$ were uniformly distributed in $(-\pi/2,\pi/2)$ and $[-\sqrt{2},\sqrt{2}]$, correspondingly. Lines which pass through the imaged square necessarily intersect two detector arrays, and one of the two intersection points was randomly chosen as the site of detection of the particle. Finally, for each such particle we recorded the bin on the detector array and the exact incoming direction $\alpha$ (see Fig~\ref{F:setup}). We also simulated an isotropic point source which emitted a number of particles roughly equal to $0.1\%$ of the background.
In order to use the filtered backprojection (FBP) inversion formula (\ref{E:FBP}) we first computed the normal coordinates of the (straight-line) trajectories of the particles:
\[\theta = \alpha-\pi/2,\quad s = (x_i,y_i)\cdot(\cos\theta,\sin\theta), \]
where $(x_i,y_i)$ is the center of the bin on the detector array. After that, the coordinates $\theta$ and $s$ were put into bins of size $0.02$. A two dimensional data array was constructed in which every element equaled the number of particles with trajectories in the corresponding $\theta, s$ bin. Different tomographic inversions were applied to such data.  The criterion for successful detection was whether the highest pick of the reconstructed image occurred in a small neighbourhood of the source location.

In our first experiment, we simulate about $300,000$ background particles and additional $300$ particles coming from a source located at $(0.401,-0.133)$. In most cases, filtered backprojection reconstruction (\ref{E:FBP}) would fail to detect the source: only $7$ out of $20$ experiments were successful. However, increasing the total number of detected particles, while keeping the SNR at $0.1\%$, improved the results significantly. Setting the background to $400,000$ and then to $500,000$ particles resulted in $13$ and $18$ detections out of $20$ experiments, correspondingly.

As we have already mentioned, the \textbf{local tomography} (which uses a stronger filter, see (\ref{E:loc})) emphasizes the singularities of an image, and thus one could expect it to detect geometrically small sources better. This, however, turns out not to be the case. In our experiments local tomography always showed inferior results to filtered backprojection, often failing to detect a source when an FBP detection from the same data was successful. A stronger high-pass filter just increases the contribution from the noise, so the reconstructed images, in fact, worsened. This suggests that maybe one should try to eliminate the filtration step altogether and relay upon the backprojection alone.

\begin{figure}[!ht]
\begin{center}
\scalebox{0.32}{\includegraphics{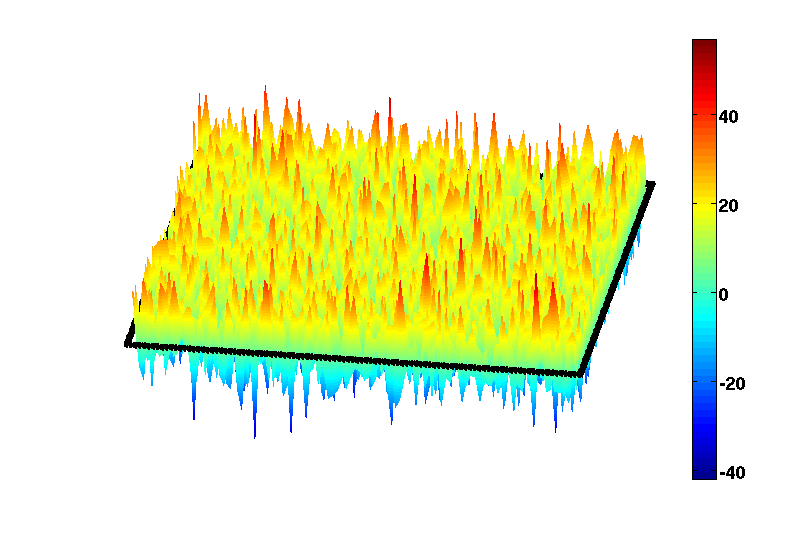}}
\scalebox{0.32}{\includegraphics{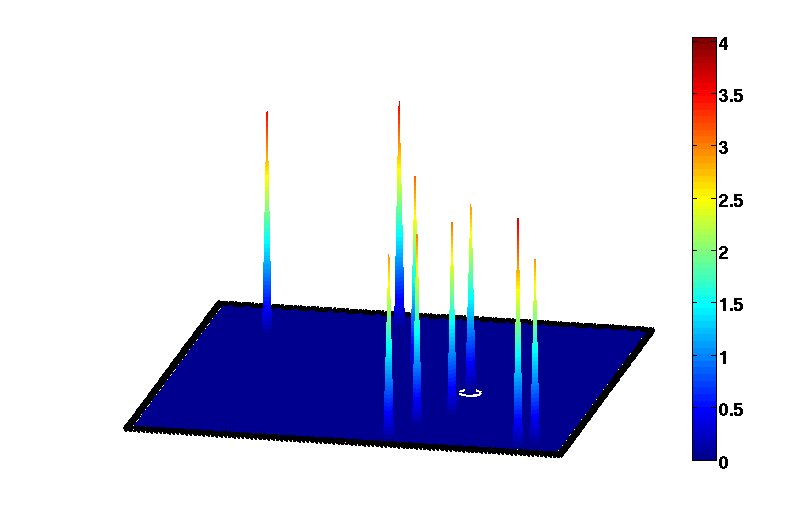}}\\
\scalebox{0.32}{\includegraphics{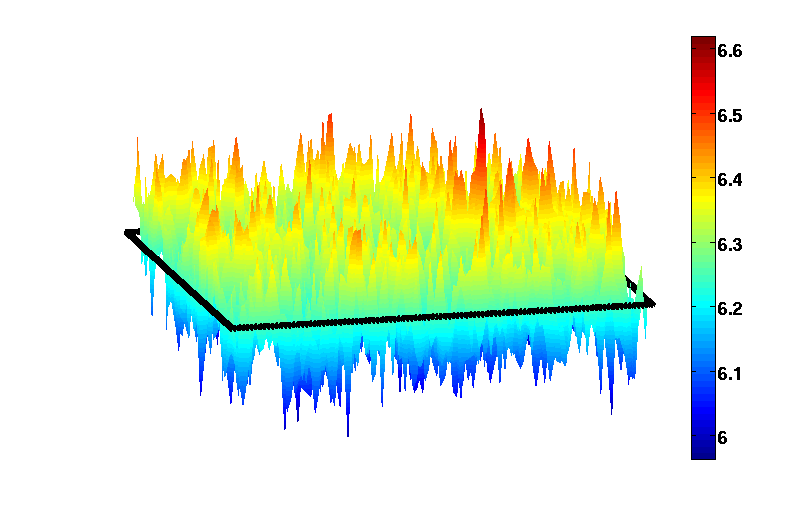}}
\scalebox{0.32}{\includegraphics{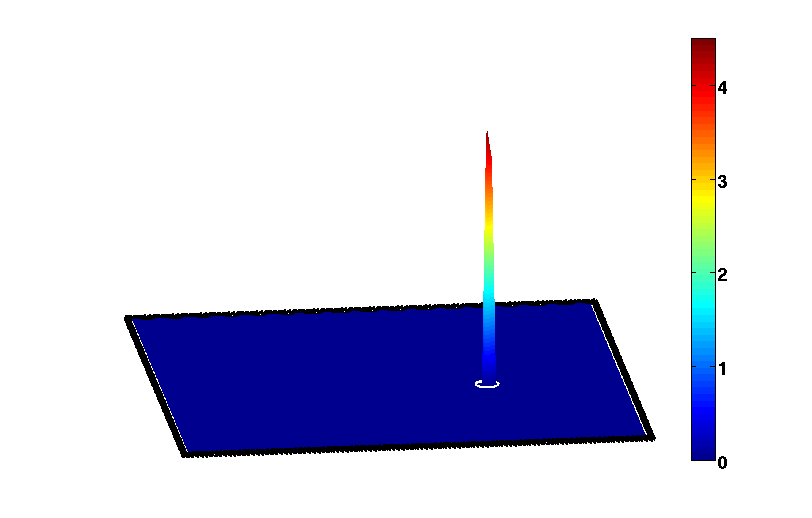}}
\caption{Reconstructions from x-ray data generated by $499,765$ random background particles and $500$ particles emitted by an isotropic source located at $(0.401,-0.133)$. The black lines indicate the location of detector arrays. The correct source location is encircled.
Top row: FBP reconstruction (left), only peaks deviating more than $3.2$ standard deviations from the mean (right). A number of peaks in the image are higher than the peak at the source.
Bottom row: backprojection reconstruction (left), only peaks deviating more than $3.5$ standard deviations from the mean (right). The correct location of the source is found.}
\label{F:xray_fbp_vs_bp}
\end{center}
\end{figure}

Indeed, our results using backprojection with no filtration were very encouraging. Backprojection of the same data used for FBP and local reconstructions resulted in $16, 18$ and $20$ out of $20$ successful detections for $300,000$, $400,000$ and $500,000$ detected background particles, correspondingly. The results of these experiments are summarized in Table~\ref{T:tomorec}.
Fig~\ref{F:xray_fbp_vs_bp} shows an example of an unsuccessful FBP reconstruction and the corresponding successful backprojection reconstruction from data generated by $499,765$ background particles and $500$ source paricles.
\begin{table}[!htb]
\label{T:tomorec}
\begin{center}
 \begin{tabular}{|c|c|c|c|}\hline
 method $\backslash$ bkgd particles & $300,000$ & $400,000$ & $500,000$  \\\hline
  Backprojection    		 & $16/20$   & $18/20$   &  $20/20$  \\
  FBP               	 	 &  $7/20$   & $13/20$   &  $18/20$  \\
  Local tomography  		 &  $6/20$   & $9/20$   &   $13/20$  \\\hline
\end{tabular}\\
\caption{Number of successful detections of a source located at $(0.401,-0.133)$, with a fixed SNR $0.1\%$ and varying number of detected particles, using different reconstruction methods. For each level of background particles $20$ sets of random data were generated. In each case backprojection, FBP and local tomography were used to detect the source. In all experiments, if a local tomography detection was successful, so was the corresponding FBP detection, and if there was an FBP detection, backprojection would also find the source.}
\end{center}
\end{table}

\subsection{$3D$ X-ray and Radon reconstructions}\label{SS:3DX}

While in $2D$ there is a single transform that can be called interchangeably Radon transform or X-ray transform, these notions part in higher dimensions. In $3D$, there is the \textbf{X-ray transform}, which \textbf{integrates a function $f(x)$ over all straight lines} in $\RR^3$, as well as the \textbf{Radon transform}, which \textbf{integrates functions over all $2D$ planes}. Both are of interest in tomography, the first arising in X-ray CT, SPECT, and PET scanners and the second in MRI imaging. In the situation we consider in this text, X-ray transform represents, the same way as it does in $2D$, the case of collimated detectors, while the Radon transform will have some relation to the Compton camera case considered in Section \ref{S:compton}. Indeed, in both $3D$ Radon and Compton cases, the data represents surface integrals (over planes and cones respectively).

\subsubsection{$3D$ X-ray reconstructions}

The same issues that we have discussed in $2D$ case remain: the X-ray transform model is valid only for strong sources (or long observation time) and it does not address the background noise. However, in $3D$ there is another feature that one has to take into account. Namely, while in $2D$ the set of lines is also $2$-dimensional (the lines can be parametrized by their two normal coordinates), in $3D$ the space of lines is $4$-dimensional. Thus, the X-ray data in $3D$ is redundant. In tomography, this redundancy is usually dealt with by selecting a $3$-dimensional sub-set of data (the rest might not even been collected). For instance, one can use only rays parallel to a fixed plane and do the reconstruction slice-by-slice, or one can pick only lines intersecting a given curve (often a helix). In the detection situation we are facing, this is not an option, since any attempt to reduce dimension of the data will essentially  remove all signal and leave only noise in its wake. Thus, we need to use all, overdetermined data. This requires corresponding analytic formulas (which are easy to obtain and thus will not be introduced) and much heavier calculations.

\begin{figure}[ht!]
\begin{center}
\scalebox{0.2}{\includegraphics{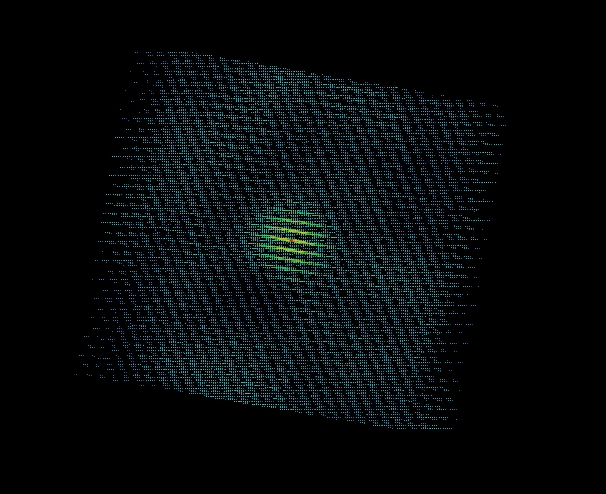}}\\
\scalebox{0.25}{\includegraphics{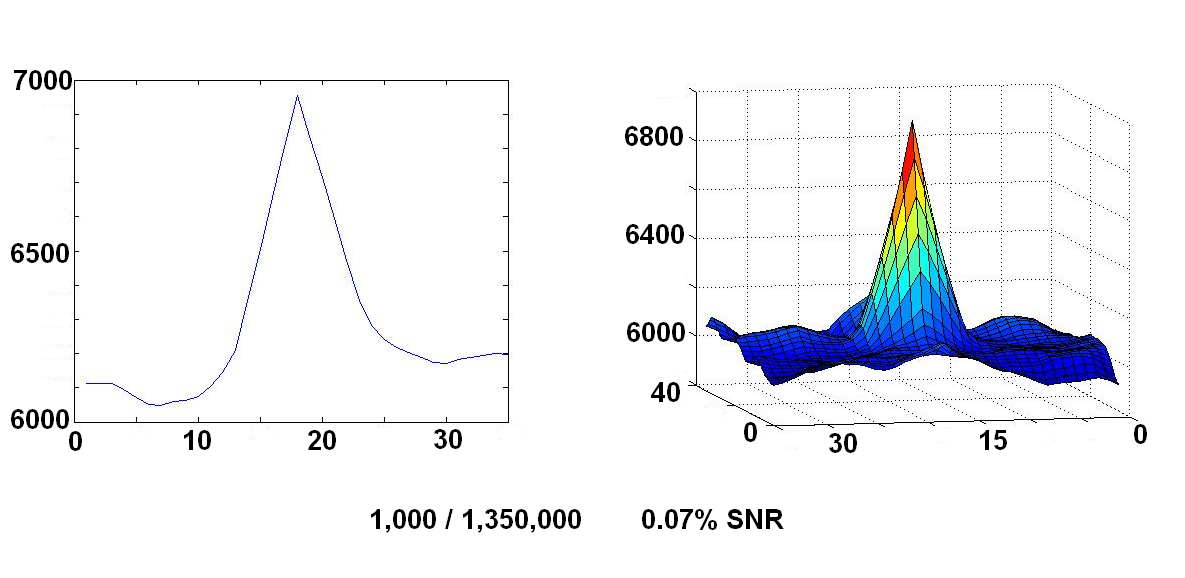}}
\caption{
Here one sees a ``transparent'' $3D$ visualization, as well as pieces of $1D$ and $2D$ sections passing through the source.}
\label{F:3d1_1345}
\end{center}
\end{figure}

Without further ado, we present an example (Fig. \ref{F:3d1_1345}) of a successful FBP reconstruction, where $10^3$ ballistic particles from the source and $1.35\times 10^6$ background ones were detected.

Further experimenting with examples leads to the same conclusion as in $2D$: backprojection is superior to FBP in this situation, while local reconstruction leads to deterioration.

\subsubsection{$3D$ Radon reconstructions}\label{SS:3DR}

In this section, we present some Radon transform reconstructions in $3D$. The reader should recall that the $3D$ Radon transform integrates functions over affine $2D$ planes, rather than lines. This is a preliminary test of the principle before doing $3D$ reconstructions for Compton cameras, where surface integrals (over cones) are also involved.

In the figures \ref{F:5e3} -- \ref{F:7e2} below, there were $10^6$ background particles detected, while the number of ballistic particles from the source was varying, thus changing the SNR and the level of detectability. Density and surface plots of $2D$ sections through the source locations are shown.
\begin{figure}[ht!]
\begin{center}
    \scalebox{0.19}{
      \includegraphics{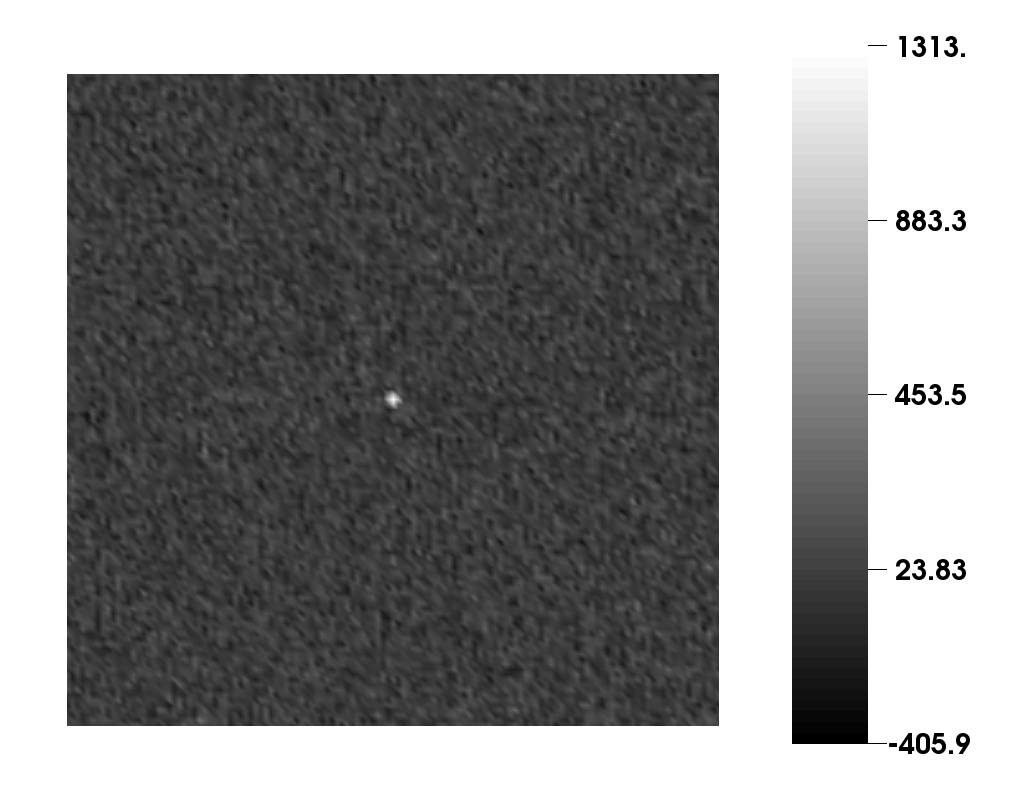}
    	\includegraphics{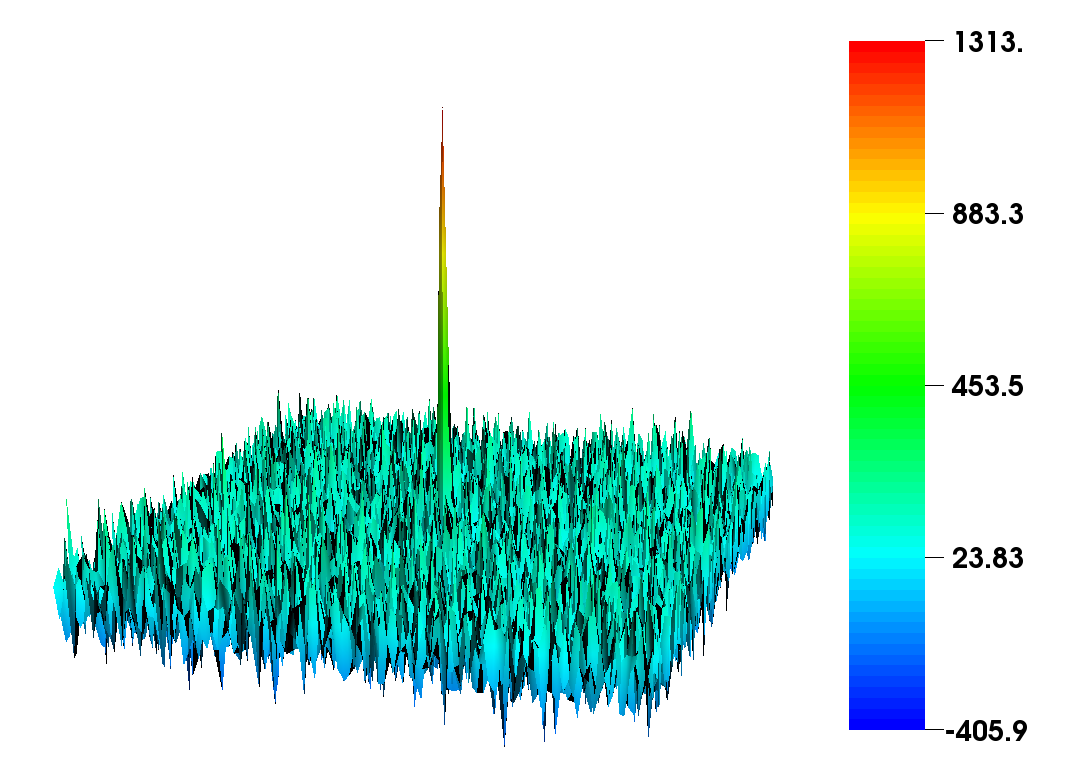}
    } \\
    \caption{FBP reconstruction from $3D$ Radon data. $5000$ ballistic source particles. SNR $0.5\%$.}\label{F:5e3}
\end{center}
\end{figure}
\begin{figure}[ht!]
\begin{center}
    \scalebox{0.19}{
      \includegraphics{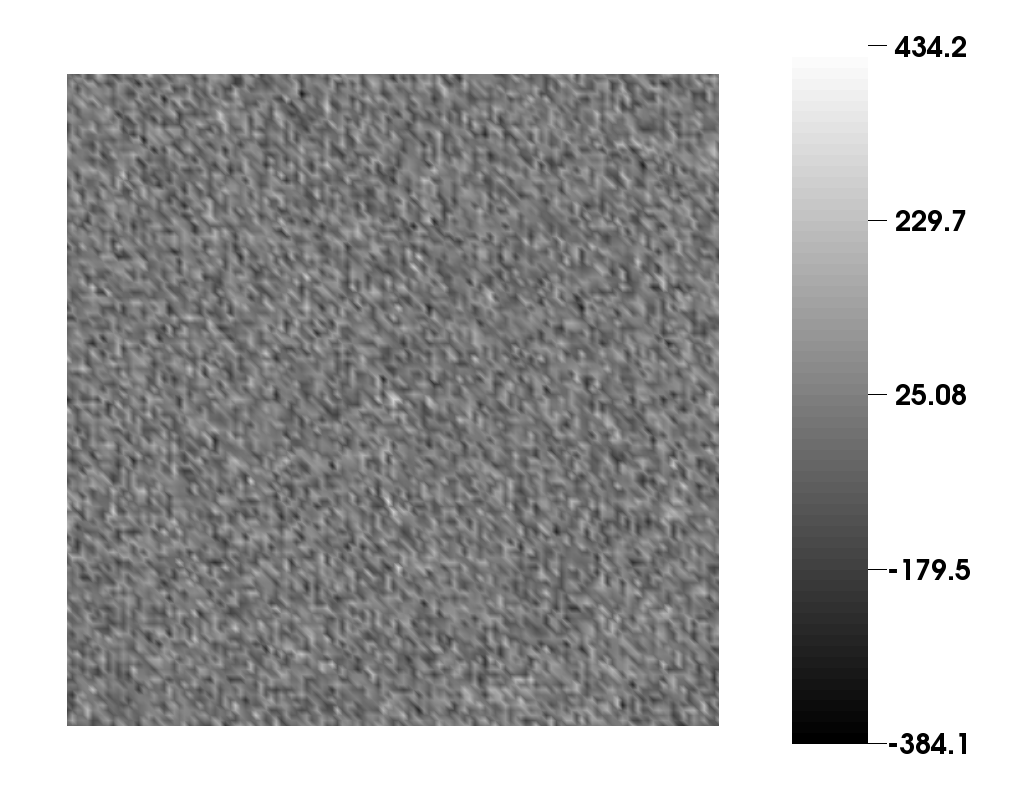}
    	\includegraphics{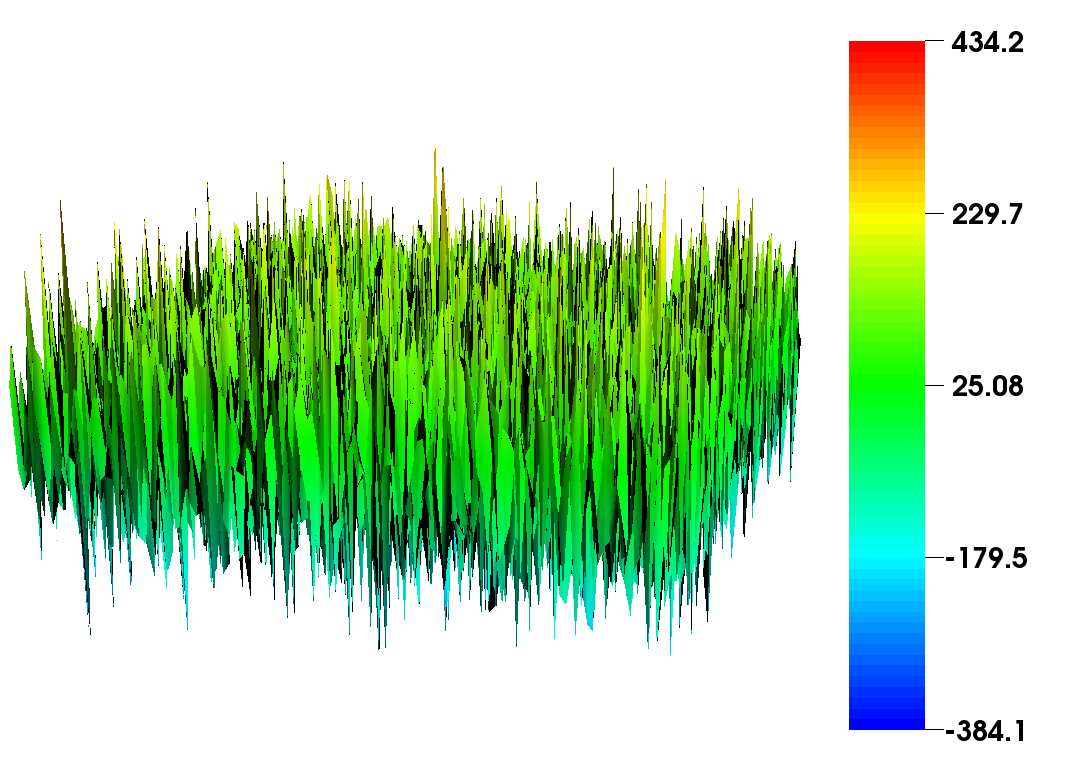}		
    }    \\
    \caption{FBP reconstruction from $3D$ Radon data. $1000$ ballistic source particles. SNR $0.1\%$.}\label{F:1e3}
\end{center}
\end{figure}
\begin{figure}[ht!]
\begin{center}
    \scalebox{0.19}{
    	\includegraphics{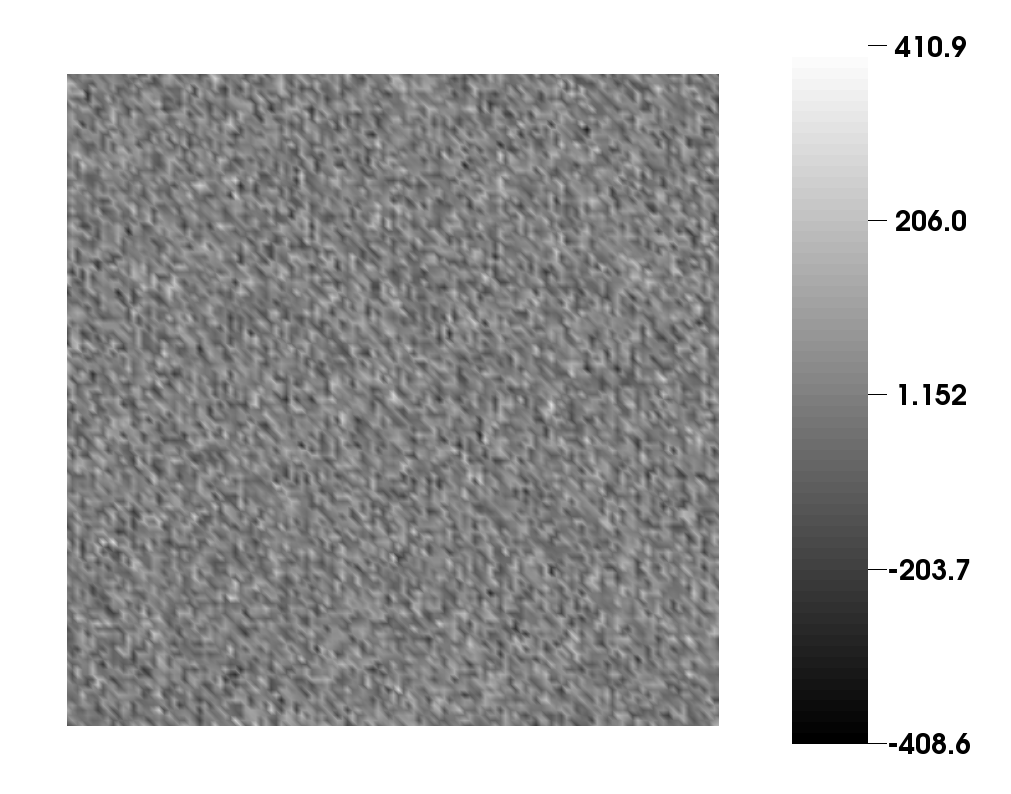}
    	\includegraphics{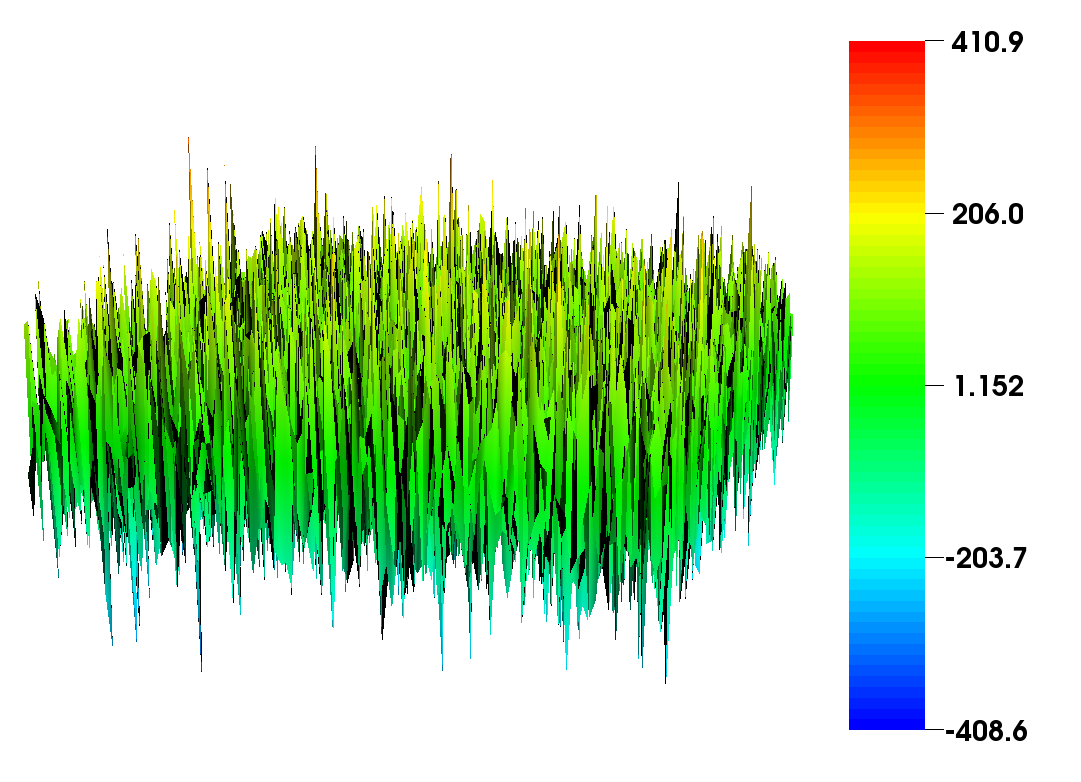}
    }    \\
    \caption{FBP reconstruction from $3D$ Radon data. $700$ ballistic source particles. SNR $0.07\%$.}\label{F:7e2}
\end{center}
\end{figure}
One sees that with the SNR of $0.5\%$ the source can be detected, while at  $0.1\%$ and lower the detection deteriorates.

\begin{figure}[ht!]
\begin{center}
    \scalebox{0.19}{
			\includegraphics{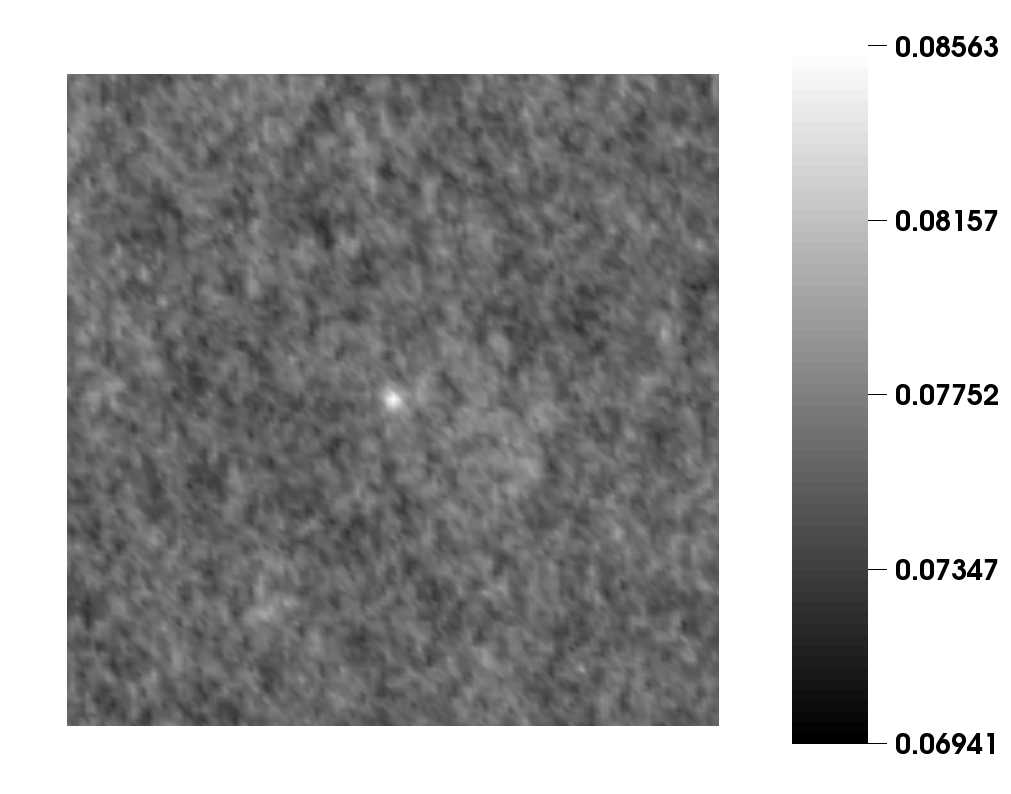}
			\includegraphics{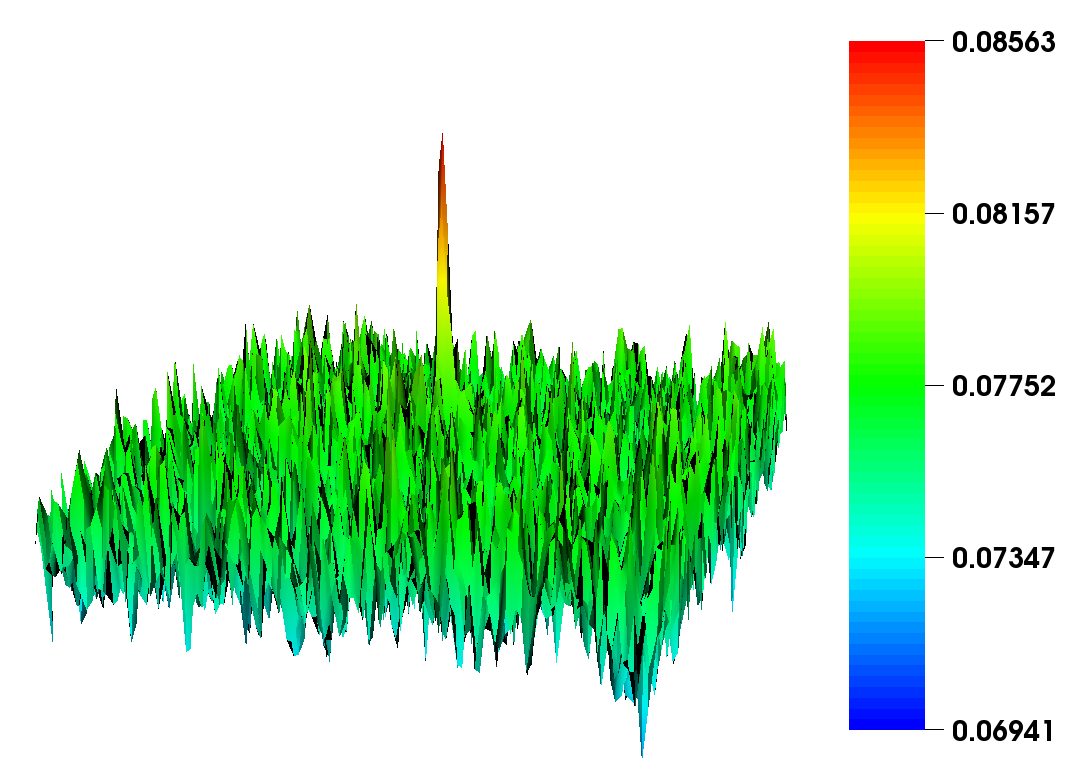}}
			
			\scalebox{0.19}{
			\includegraphics{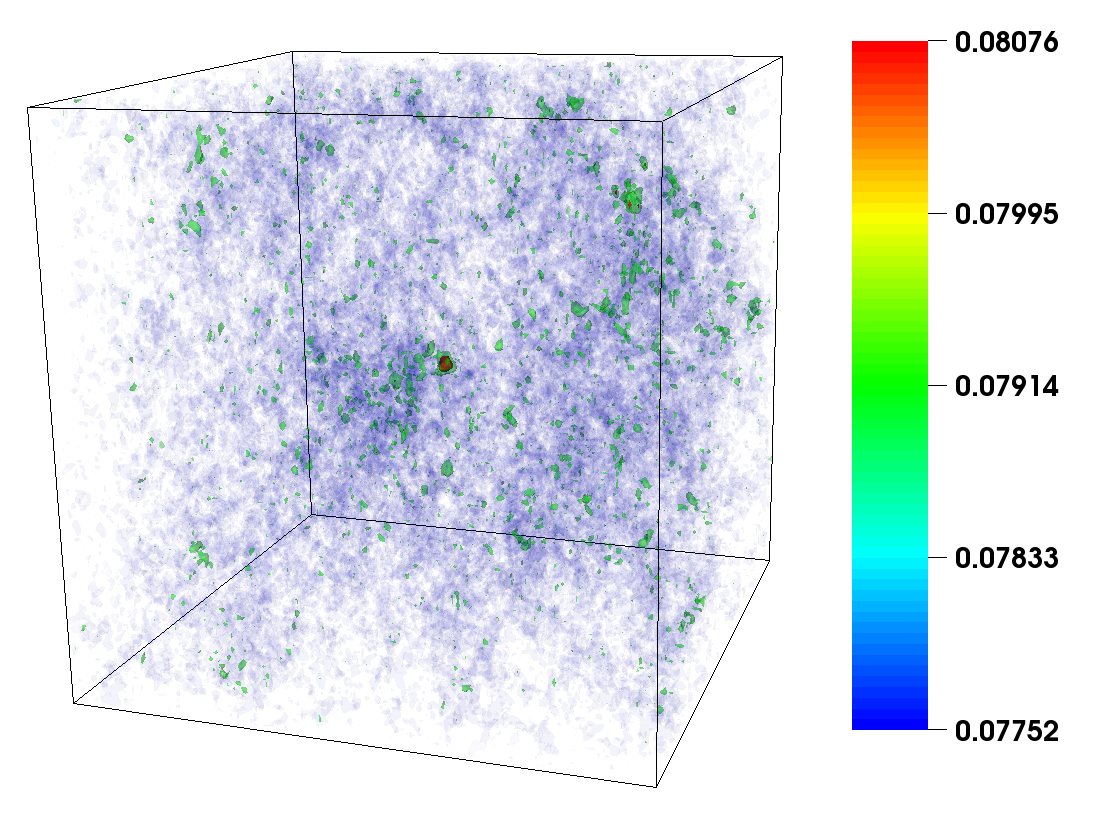}}
    \caption{BP reconstruction from $3D$ Radon data. $1000$ ballistic source
particles, SNR $0.1\%$. Top: Cross section through source. Bottom: Isosurfaces
at 50\% (blue), 60\% (green) and 70\% (red) of peak value.}\label{F:bp1e3}
\end{center}
\end{figure}

\begin{figure}[ht!]
\begin{center}
    \scalebox{0.19}{
    \includegraphics{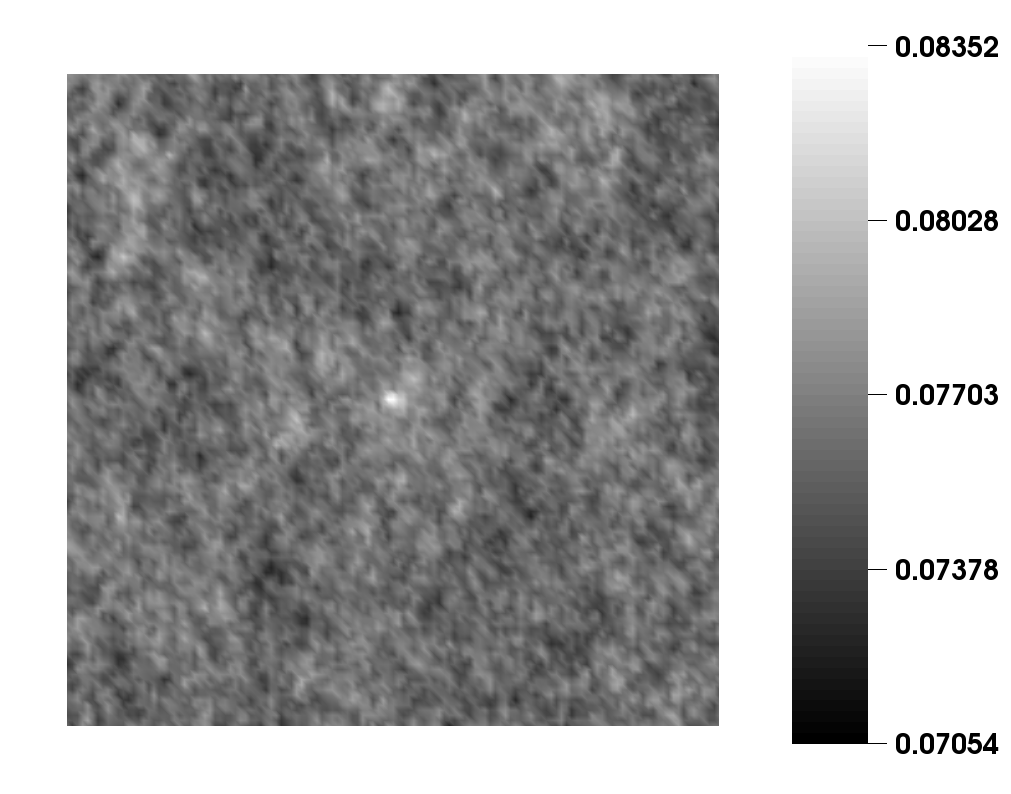}
    \includegraphics{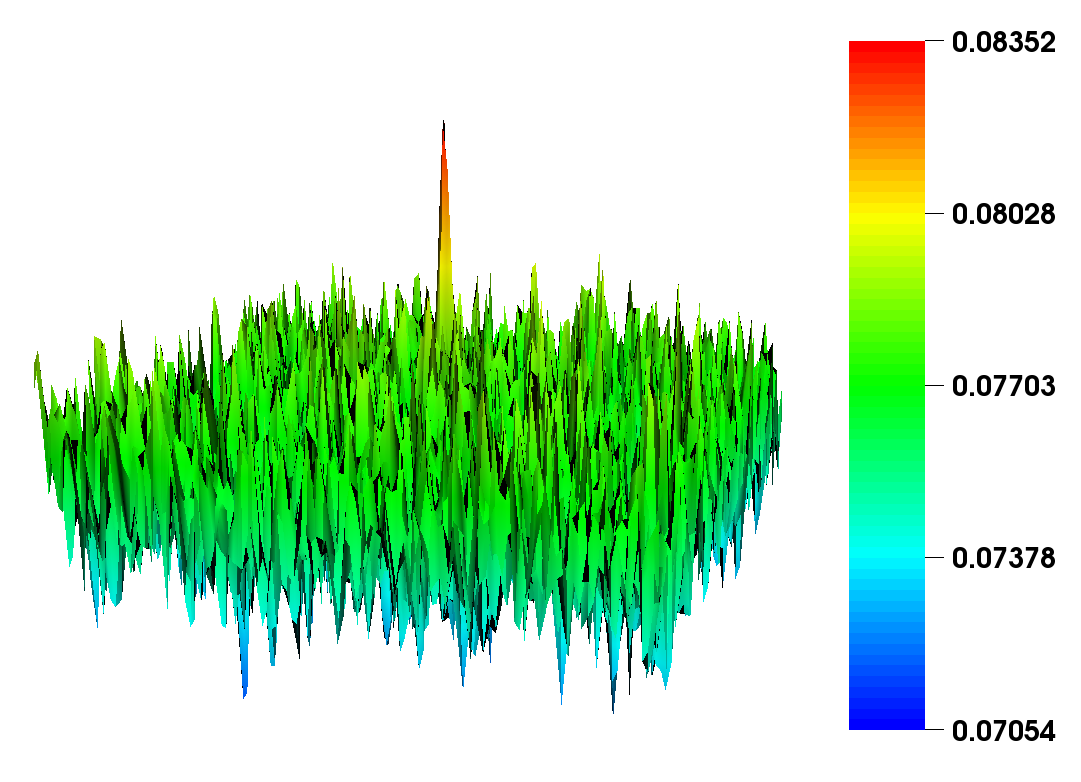}}

    \scalebox{0.19}{
    \includegraphics{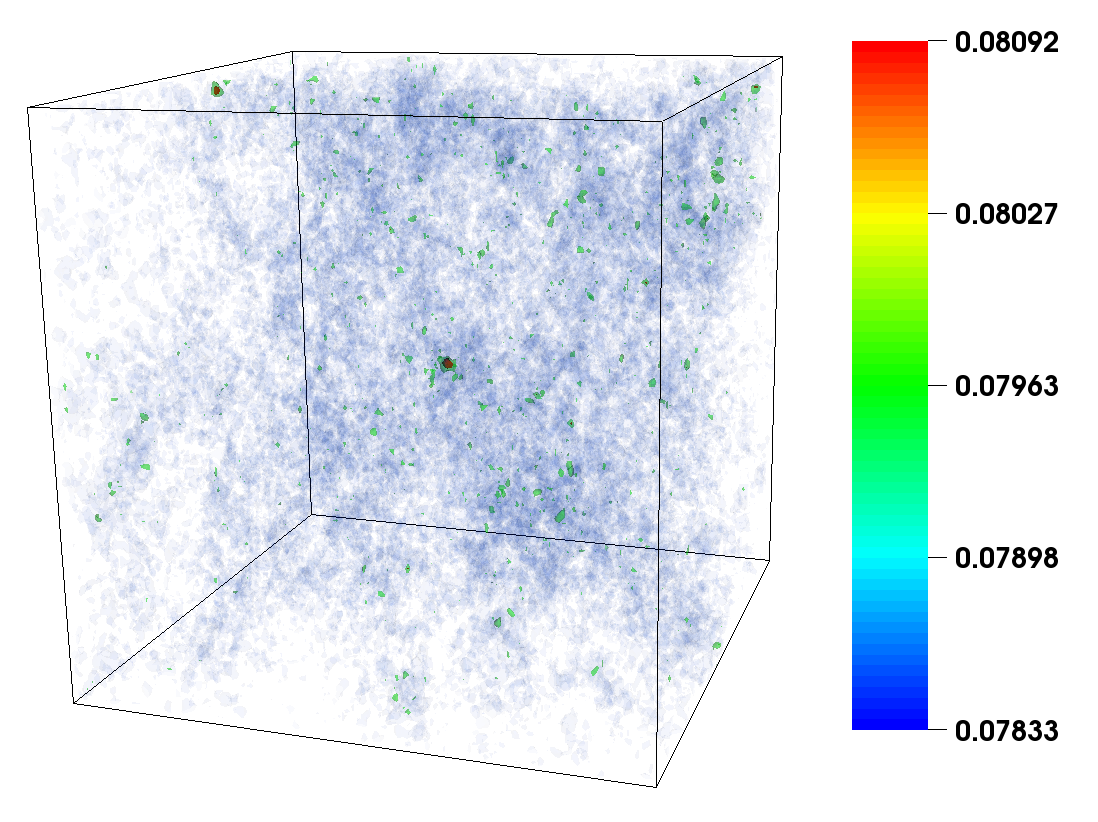}}
    \caption{BP reconstruction from $3D$ Radon data. $700$ ballistic source
particles, SNR $0.07\%$. Top: Cross section through source. Bottom: Isosurfaces
at 55\% (blue), 70\% (green) and 80\% (red) of peak value.}\label{F:bp7e2}
\end{center}
\end{figure}

Let us see now whether backprojection alone does a better job. Figures \ref{F:bp1e3} and \ref{F:bp7e2} show the backprojection reconstruction from the same data as in Fig. \ref{F:1e3} and \ref{F:7e2}. One clearly sees that the backprojection alone detects the source when the FBP method fails to do so.

Again, the examples above confirm the general feasibility of the approach and superiority of the backprojection.

\subsection{Discussion of the tomographic reconstructions}\label{SS:discussion}

The experiments described in the previous sections beg several questions:
\begin{enumerate}
  \item Why do tomographic methods still work to some degree, although the assumptions that lead to the X-ray transform model are not satisfied (i.e., the source is too weak) and a strong random background is present? Indeed, the use
of standard tomography assumes that the data represents line integrals of the source
distribution function. This is the same as to say that we measure the expectation of the number of
hits, per unit time, from particles moving along a line. This assumption is reasonable
when a substantial number of source particles is detected, i.e. when the source is strong, or the observation time is long. However, it fails in the case of low emission sources that cannot be observed for a really long time. Hence, the X-ray/Radon transform type models and techniques appear to be inappropriate.
  \item Why does strengthening the filter (i.e., local tomography), which should increase detectability of singularities, makes the reconstruction worse?
  \item Why do tomographic techniques fail to reliably detect sources at the levels of SNR and total number of  particles for which the probabilistic consideration of the next section suggests that sources should be detectable?
\end{enumerate}

As we might have guessed already, the answer to the first question is that in fact, tomographic inversions  do not work. Or, to put it differently, only a part of the algorithm (backprojection) works, while its other part (filtration) only hurts.

The answer to the second question is not hard to guess. Indeed, high-pass filters required
in X-ray/Radon transform inversions amplify noise. This is a very significant factor, as the noise
constitutes $99.9\%$ of our data. Using local tomography can only aggravate this difficulty, and we have seen that it indeed does.

The answer to the third question is probably the same as the previous one: high-pass filters might be the culprits.

These considerations suggest to try to eliminate high-pass filtering completely, and thus to use backprojection alone (not a good idea in the standard imaging, where it leads to blurring \cite{Kak,Natterer_book}). We will see in the next sections that this is indeed the solution. Simultaneously, we will eliminate another drawback of the standard tomographic techniques, which is the difficulty of quantifying the confidence level of detection.

\section{A probabilistic discussion}\label{S:prob}

Now we investigate why it \emph{might} be possible to detect \emph{geometrically small} low emission sources by backprojection.
Let us suppose that the detectors are collimated and thus can determine information about the incident direction of particles. Then backprojection essentially ``sends'' the detected particles back along their incoming straight lines. In particular, these backpropagated trajectories of all ballistic source particles will pass through the source region. The background particles or the source particles that have scattered before reaching the detector, will be backprojected along straight lines, which are different from their original zig-zag trajectories, about which we have no knowledge. Therefore, in the considerations that follow, we  assume that the backpropagated trajectories of these particles are random straight lines. We also assume for now that the random distribution of trajectories is uniform.

\begin{figure}[!ht]
\begin{center}
\scalebox{0.5}{\includegraphics{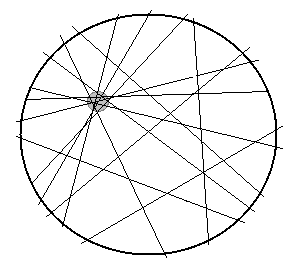}}
\caption{If the local density of lines is significantly higher in the gray region, this is likely due to a source being present at this location. The large circle indicates the region of interest (ROI) rather than any physical boundary.}
\label{F:CMP_lines}
\end{center}
\end{figure}

In a nutshell, our simple argument, which will be made more precise below, is that \emph{if the number of lines passing through a tiny region exceeds significantly the mean of the background, then most probably there is a source at this location}
(see Fig. \ref{F:CMP_lines}).

Let us try to quantify this and consider $N$ random particle trajectories in a ball $B_R$ of radius $R$. What is the probability of at least $n$ particles out of the total $N$ contained in $B_R$ to pass through a small
ball $B_r$ with radius $r$? Let, as before, $L_{(\omega,s)}$ denote the line in $\RR^2$ defined by its unit normal vector
$\momega \in S^1$ and a signed distance $s$ from the origin, i.e.
$$
L_{(\omega,s)}=\{\mx\in\RR^2|\mx\cdot\momega=s\}.
$$
Thus, the lines on the plane are mapped 1-to-1 to the points on the
half-cylinder $S_+^1\times (-\infty, \infty)$, where $S_+^1 = \{\momega = (\cos\theta,\sin\theta),
\theta\in[0,\pi) \}$, and these points are uniformly distributed on the half-cylinder.  All lines
intersecting $B_R$ correspond to points on $S_+^1\times(-R,R)$, and those intersecting $B_r$ are
bijectively mapped to points on $S_+^1\times(-r,r)$. Thus, the probability of a random line in $B_R$
passing through $B_r$ is
\begin{equation*}
p=\frac{\mbox{area}(S_+^1\times(-r,r))}{\mbox{area}(S_+^1\times(-R,R))} = \frac{r}{R}.
\end{equation*}
The probability that $n$ out of the total $N$ lines cross $B_r$ is given by the binomial
distribution $B(N,p)$, and is equal to $\displaystyle\binom {N}{n} p^n(1-p)^{N-n}$. Since we are in
the situation when $p$ is fixed by the dimension of the source and $N$ is large, the Central Limit
Theorem applies. It follows (since the values of $Np$ we will see will range in thousands) that the binomial distribution is approximated well by the Normal
distribution $N(\mu,\sigma^2)$ with the mean $\mu = Np$ and standard deviation $\sigma=\sqrt{Np(1-p)}$.
If it happens that the number of lines $n$ crossing $B_r$ exceeds significantly (in comparison with
$\sigma$) the mean $\mu$, the probability of this occurring just due to random reasons is very
small. Therefore, one can be almost certain that this clustering of trajectories is the result of a
radioactive source located at $B_r$.
Our numerical experiments of source detection by backprojection agree with these expectations.

\begin{figure}[!ht]
\begin{center}
\scalebox{0.35}{\includegraphics{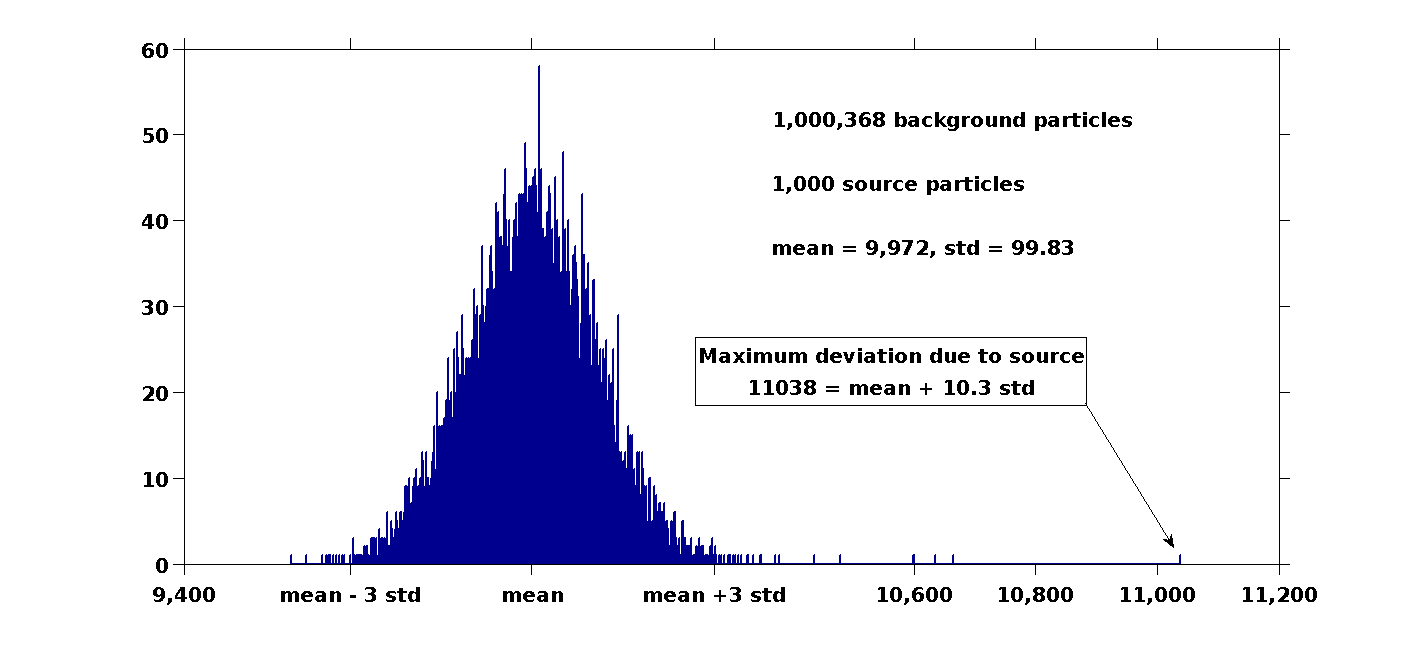}}
\caption{Histogram of a backprojection reconstruction from data collected by four collimated detectors, placed on the sides of the unit square. The data simulated a source at the point $(0.3,-0.4)$, with $1,000,368$ background particles and $1,000$ source particles. The mean of the backprojected image is $9,972$, and the standard deviation is $99.83$. The number of lines at the location of the source deviates $10.3$ standard deviations from the mean. The pixels in close proximity of the source also have an elevated number of lines passing through them and thus are responsible for several other far to the right outliers. They would show as a single pick on the detection plot.}
\label{F:CMP_deviation}
\end{center}
\end{figure}

As an example, in Fig.~\ref{F:CMP_deviation} we present a typical histogram of the number of lines crossing a pixel in the square $D = [-1,1]^2$ with a uniform grid of $100^2$ pixels. As described in section \ref{SS:2DX}, we generated random background and an isotropic point source. The particles were detected by four arrays of collimated detectors placed on the sides of the square. The exact incoming directions of particles were recorded. For the result shown below, we backprojected each particle using a line-drawing algorithm. Namely, each particle was back propagated along a straight line into $D$ by adding a value of $1$ to all pixels in $D$ intersected by the line. After all particles were backprojected, the value at each pixel of $D$ is equal to the number of lines intersecting it. On figure \ref{F:CMP_deviation} we show a histogram of the pixel values after bakprojecting particles coming from a random background of about $10^6$ particles and $10^3$ source particles. On the far right of the histogram one sees the outlier contribution from the source, which deviates about $10.3$ standard deviations from the mean\footnote{There are several other outliers, deviating by 4 -- 8 standard deviations. They all correspond to the pixels intersecting the same source.}. This result complies with discussion above: the appearance of such a highly unlikely outlier indicates presence of a source.

Now we will make these statements more precise.
Suppose that the area we are imaging is divided into $N_{pix}$ pixels. Suppose further that $n$
lines intersect a pixel $B_r$ with radius $r$. We will write $n = n_s+n_b$, where $n_s$ and $n_b$
are the number of ballistic source and background trajectories, respectively, which cross $B_r$.  We will
describe a method to determine whether there is a source located at this pixel.

Let us choose a threshold value $k_t$, such that the probability of a normal variable to reach more than $k_t$ standard deviations above the mean is very small. E.g., $k_t=5$ or higher suffices. If an abnormally high number $n$ of lines passing through a pixel $B_r$ is detected, i.e. if $n>n_t:=\mu+k_t\sigma$, we will claim that the pixel contains a source. Otherwise, such claim will not be made. The probability of at least $n_t$ lines crossing $B_r$ due to random reasons is approximately
$r:=0.5\;\mbox{erfc}(k_t/\sqrt{2})$. We also have to take into account the total number $N_{pix}$ of pixels and the possibility of such clustering
of lines occurring inside at least one of them. The probability of at least $n_t$
lines crossing at least one of the pixels due to random reasons (and thus causing a "false alarm") can be approximated by
\begin{equation}\label{E:FP_rate}
 \mbox{fp rate} = 1-(1-r)^{N_{pix}}.
\end{equation}
Here \emph{fp rate} stands for \emph{false positive rate}, which is the rate of false detections
\cite{Fawcett_ROC}. The above estimate assumed independence across different pixels of the events ``the pixel is crossed by $n$ lines''. It is easy to see that this is in fact not the case: If a pixel is crossed by a large number of lines, there is a good chance that the surrounding pixels would also have a many lines intersecting. However, we claim that treating them as independent does not have a significant impact on the resulting estimates. To confirm this, we show some statistics from numerical simulations of random backgrounds at the end of this section.

The \emph{true negative rate} (or specificity), is given by $$
\mbox{tn rate} = 1-\mbox{fp rate}
$$
and
represents the rate of correctly classified negative outcomes. The true negative rate is also the
confidence probability that we have found a true source. We will write
\begin{equation}\label{E:CMP_conf}
\mbox{tn rate} = \mbox{confidence} := (1-r)^{N_{pix}} = [1-0.5\mbox{ erfc}(k_t/\sqrt{2})]^{N_{pix}}.
\end{equation}
In particular, if the threshold value $k_t$ is set to $5$ and there are $100^2$ pixels, the true negative rate is $.997$, which gives also this number as the confidence that the alleged sources are indeed true.
This confidence probability depends only on the pre-set threshold value $k_t$ of the method and the number of
pixels $N_{pix}$. However, if in some pixel the number of lines deviates from the mean even further, this increases the confidence probability that there is indeed a source there. For instance, if $n\geq\mu+7\sigma$, the confidence reaches, for all practical purposes, $100\%$.

Let us discuss now the probability of missing a source.
We assume that an \textit{ apriori} ballpark estimate of possible value of $n_s$ (the number of ballistic particles coming from the source) is available. We can measure $n_s$ in the $\sigma$ units:
\[
n_s = k_s\sigma.
\]
Here $k_s$ is a positive constant and $\sigma$ is the standard deviation of the distribution of the number of background trajectories passing through a pixel\footnote{
Recall that $\sigma$ is determined by the total number of detected particles, which is known, and
the ratios of the source dimension to the dimension of the imaged region, for which we would, in
practice, have a crude idea.}.
Now suppose that there is indeed a source present in a pixel $B_r$. How can we miss it? Since our test identifies a source whenever $n=n_s+n_b>\mu+k_t\sigma$, the necessary condition for missing it is
$n_b<\mu-(k_s-k_t)\sigma.$
The probability of this happening (i.e., the \emph{false negative rate} (fn rate)) is given by
\begin{equation}\label{E:CMP_fnrate}
 \mbox{fn rate} = 0.5\mbox{ erfc}\left(\frac{k_{s}-k_t}{\sqrt{2}}\right).
\end{equation}
For instance, if $k_s\geq k_t+3$, the probability of missing this source is at most $0.0013$.

The
sensitivity, or \emph{true positive rate} (tp rate), is given by
\[
\mbox{tp rate} = 1-\mbox{fn rate}.
\]

Note that the sensitivity of the method depends on the threshold value $k_t$ as well as on our
estimate of detected source particles.

From the above discussion we see that in order to find sources reliably, it suffices to detect $n_s > (k_t+3)\sigma$ ballistic particles from the source.

We claim that \textbf{for any given SNR $s>0$ and for any $k_t>0$ it is possible to detect the source, provided that the total number of detected particles, $N$, is sufficiently large.}

Indeed, if a signal to noise ratio $s$ is known, and if $p\ll 1$, the number of detected source particles will be
$n_s\approx sN$ and $\sigma\approx \sqrt{Np(1-p)}$. Then, we require the following inequality to be satisfied:
\begin{equation}\label{E:CMP_prob}
sN > (k_t+3)\sqrt{Np(1-p)}.
\end{equation}
Since the left hand side of (\ref{E:CMP_prob}) grows linearly with $N$ and the right hand side
grows only as $\sqrt{N}$, the inequality will be satisfied for a sufficiently large $N$. We come up with the following rule of thumb:

\textbf{One can expect to detect reliably the source with high sensitivity and specificity, if the total particle count $N$ at the detectors is on the order of
\begin{equation}\label{E:N}
N\propto\left(\frac{8}{s}\right)^2p(1-p),
\end{equation}
or higher, where $s$ is the SNR (ballistic particles from the source vs the total hits) and $p=r/R$, where $r$ and $R$ are the linear dimensions of the source and the whole area correspondingly.}

Let us play with some practically possible values of the parameters. A nuclear source with shielding could have a radius of the order of several
centimeters, while a vehicle or a cargo container has size in the order of meters. Thus, we will
assume that
\begin{equation}\label{E:CMP_dim}
R=1, \;  r=0.01 \;\;\Rightarrow p=10^{-2}.
\end{equation}
%
Substituting this value of $p$ and $s\approx 10^{-3}$ into \ref{E:N}, one expects that when $N$ reaches close to $640,000$ and directional information at detectors is available, the sources should be reliably detectable. This squares well with the results of our previous computations with the backprojection methods.

The conclusion from this probabilistic discussion is that the simple backprojection should reliably detect sources with low SNR, as long as the total number $N$ of detected particles is high enough (see (\ref{E:N})). E.g., for the SNR of about $0.1\%$ the values of $N$ around $640,000$ should be sufficient. Besides, the confidence probability of detection (or absence of a source) can be computed.

The use of backprojection should also smooth out unstructured noise. In the next section we present numeric examples of such backprojection detections.

It is worthwhile to notice that, as long as the threshold parameter $k_t$ (equivalently, the desired confidence probability) is fixed, the whole detection procedure can be done automatically, without necessity of a human eye assessment of the reconstruction.

\paragraph{Comparison to simulation data.}
Recall that equation \eqref{E:FP_rate} and most of the following discussion assumed (incorrectly) independence for different pixels of the events ``a pixel is crossed by $n$ lines''. To confirm that the confidences we estimated under this assumption are not vastly optimistic, we simulate a large number of random background fields and compare the statistics of these samples to our theoretical estimates. Each sample consists of $10^6$ uniformly distributed random lines on a grid of 100$\times$100 pixels. For this setting the mean and standard deviation of the number of lines crossing a pixel are approximately $\mu=10,000$ and $\sigma=99.5$. For different values of the threshold parameter $k_t$ ranging from 4 to 5, we record the true negative rate $r_t$ of the sample population, which is the ratio of samples in which each pixel had less or equal than $n_t = \mu+k_t\sigma$ lines intersecting. We compare this to the estimated confidences given by \eqref{E:CMP_conf}. The results from 50,000 samples are shown in table \ref{T:sample_comp}. It can be seen that for all values of $k_t$ the true negative rate from the sample population is slightly lower but very close to the estimated confidences. We conclude from this that the simplifying independence assumption made in this section is reasonable and the estimated probabilities agree well the true probabilities.

\begin{table}[!htb]
\label{T:sample_comp}
\begin{center}
 \begin{tabular}{|c|c|c|c|}\hline
 $k_t$ & $n_t$ & confidence & $r_t$  \\\hline
            4    &    10,398  &    0.7285   &   0.7141 \\
          4.1    &    10,408  &    0.8134   &   0.8014 \\
          4.2    &    10,418  &    0.8751   &   0.8658 \\
          4.3    &    10,428  &    0.9182   &   0.9125 \\
          4.4    &    10,438  &    0.9473   &    0.9428 \\
          4.5    &    10,448  &    0.9666   &    0.9623 \\
          4.6    &    10,458  &     0.9791   &    0.9760 \\
          4.7    &    10,468  &     0.9871  &    0.9846 \\
          4.8   &     10,478  &     0.9921   &   0.9907 \\
          4.9   &     10,488  &    0.9952  &    0.9944 \\
            5   &     10,497  &    0.9971  &     0.9968 \\\hline
\end{tabular}\\
\parbox{8cm}{\caption{Comparison between estimated confidences and statistics from 50,000 negative samples}}
\end{center}
\end{table}

\section{Source Detection Using Backprojection}\label{S:backproj}

We illustrate the conclusions of the previous Section on a couple of typical examples of $2D$ X-ray data. The data was generated in the way described in Section \ref{SS:2DX}. The incoming trajectories of detected particles were backprojected using a line drawing algorithm, as was explained in Section \ref{S:prob}.
\begin{figure}[!ht]
\begin{center}
\scalebox{0.3}{\includegraphics{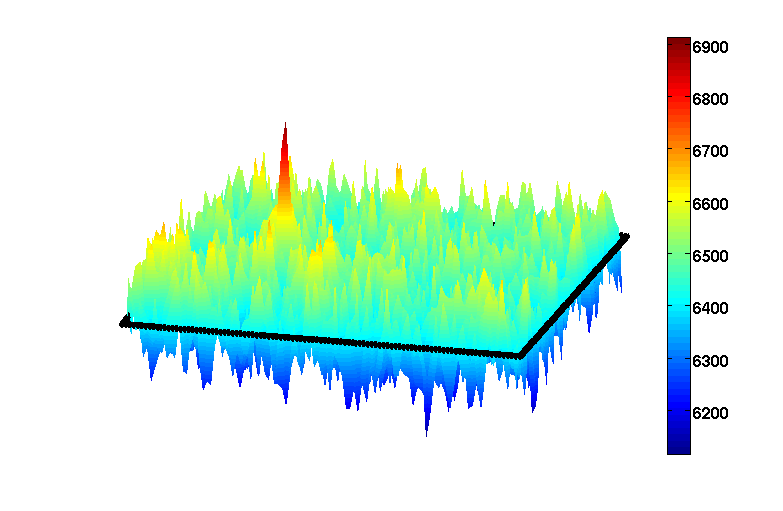}}
\scalebox{0.3}{\includegraphics{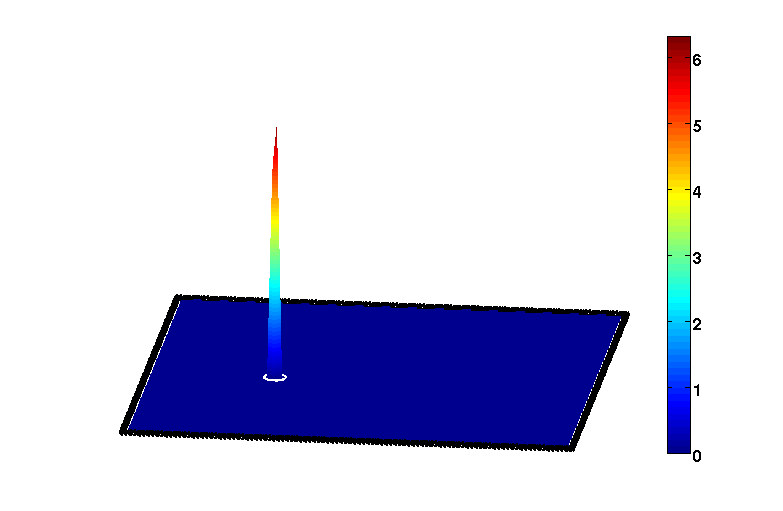}}
\caption{Left: Backprojection reconstruction from X-ray data. Right: Peaks exceeding the local mean more than $4$ standard deviations. The unit on the vertical axis is one standard deviation from the mean.
The maximum number of standard deviations above the mean equals  $6.3$ and occurs at the location of the source. The estimated confidence is $100\%$. Data consisted of $639,954$ background particles and $640$ particles coming from a source located at $(-0.43,-0.11)$. The black lines represent the location of detector arrays. }
\label{F:xray4det}
\end{center}
\end{figure}
Recall that we recorded approximate locations of the site at which a detection occurs, since in reality detector arrays consist of a number of bins (detectors).

In the first example, we simulated $639,954$ background particles and $640$ particles from a source located at $(-0.43,-0.11)$. All particles were backprojected and the source was detected with 100\% confidence (see Fig.\ref{F:xray4det}).

\begin{figure}[!ht]
\begin{center}
\scalebox{0.35}{\includegraphics{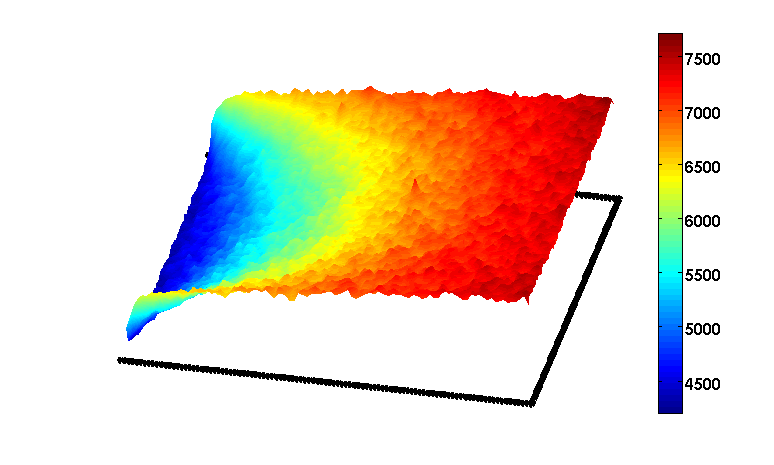}}\\
\scalebox{0.35}{\includegraphics{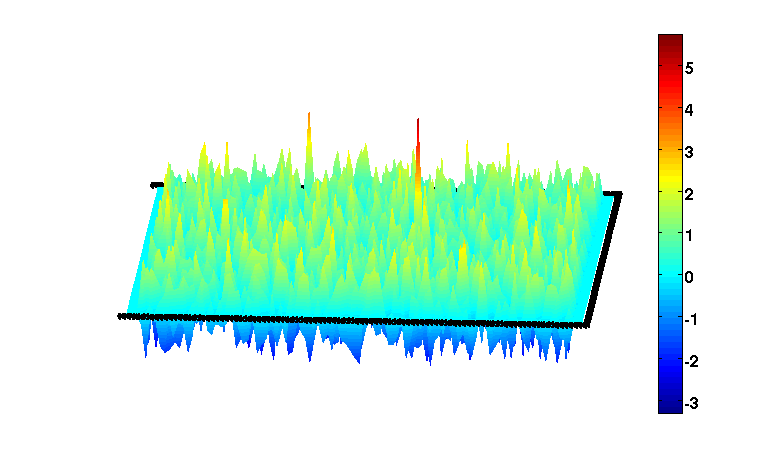}}
\scalebox{0.35}{\includegraphics{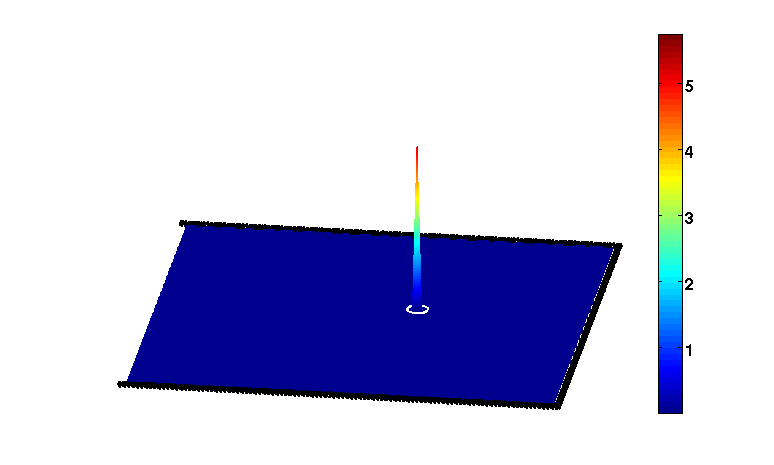}}
\caption{Backprojection reconstruction from X-ray data (top). The unit on the vertical axis in the bottom pictures is one (local) standard deviation from the (local) mean. Bottom left picture shows the result of backprojection after subtraction of the local mean. Peaks that exceed the local mean by more than $4$ (local) standard deviations are shown at bottom right. Source located at $(0.203,0.101)$ is detected with confidence $99.99\%$. Data consisted of $639,417$ background and $646$ source particles. The black lines represent the location of detector arrays.}
\label{F:CMP_RadBP}
\end{center}
\end{figure}

For our second example, we used only three detector arrays along three sides of a square. Such a ``detector gate'' is a more realistic configuration than detectors surrounding the whole object. Figure \ref{F:CMP_RadBP} shows a typical backprojection reconstruction. The drop of the values at one side of the square is due to detectors missing there. Thus, effectively, there is a non-uniform distribution of the background, which contradicts to our assumption of uniformity. However, this non-uniformity is smooth (slowly varying), so we can assume that \emph{locally} the distribution of background trajectories is uniform. In order to apply our probabilistic arguments described in Section~\ref{S:prob} we estimate the \emph{local} mean and standard deviation at every point. For the reconstruction shown on Fig.\ref{F:CMP_RadBP}, this was done by analyzing the data on a $7\times 7$ sub-grid surrounding each pixel. Then, the local means were subtracted and the resulting values plotted (see the bottom left picture in Fig.~\ref{F:CMP_RadBP}) in the units that are equal to the local standard deviation $\sigma$.
Finally, the values below the threshold $k_t=4$ were erased (bottom right in Fig.~\ref{F:CMP_RadBP}). The result clearly shows the detected source. The confidence probability was calculated as
\[
\mbox{confidence} = (1-0.5*\mbox{erfc}(k_{source}/\sqrt{2}))^{100^2},
\]
where $k_{source}$ is the height of the highest peak. The calculated confidence level for this example was $99.99\%$. Notice that if we used the threshold value $k_t=4$ instead of the actual peak's height, we would have significantly underestimated the confidence.

\section{Compton Cameras}\label{S:compton}
Compton cameras, also called Compton scatter cameras or electronically collimated cameras, have received a lot of
attention since they were first proposed in \cite{ToddNightEverett} and a prototype was developed in \cite{Singh1,Singh2}. Compton cameras were first suggested for use in nuclear medicine (SPECT),
 but they also have applications in astrophysics \cite{Schonf_Comptel,Gunter08,Watanabe}, monitoring nuclear
 power plants \cite{RoyleSpel1,RoyleSpel2}, and other areas. Compton cameras offer several advantages over
the conventional mechanically collimated (Anger) $\gamma$-cameras. The most significant of these is the dramatic
 increase of sensitivity -- Compton cameras are reported to count orders of magnitude more photons
than Anger cameras \cite{Clinthorne,Leblanc}. Another advantage is the flexibility of geometrical design
\cite{RoyleSpel1,RoyleSpel2}, which even allows for hand-held devices \cite{Lackie,HillMatt07}.

\begin{figure}[!ht]
\begin{center}
\scalebox{0.25}{\includegraphics{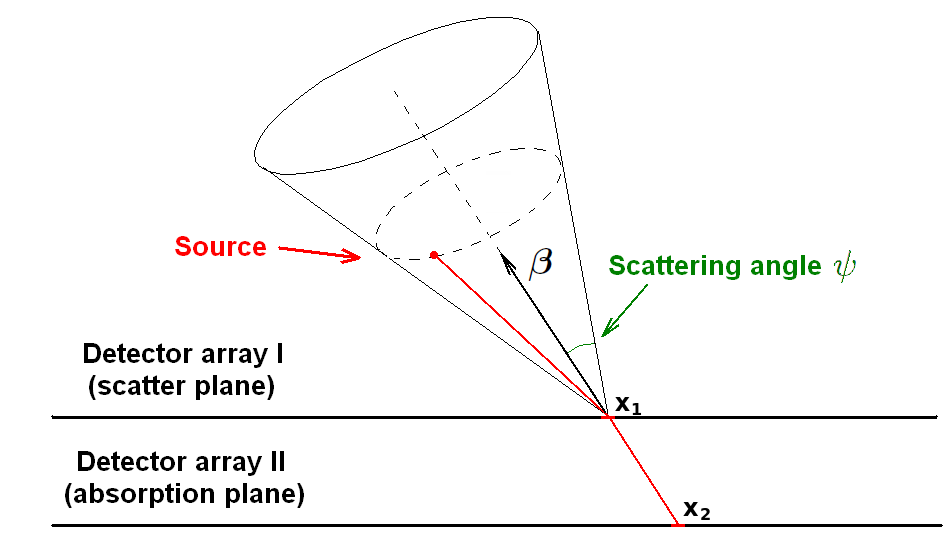}}
\caption{Schematic representation of a Compton camera.}
\label{F:compton3D}
\end{center}
\end{figure}

A Compton camera consists of two detectors, which, in the standard setup, are planar and are placed one
behind the other (see Fig. \ref{F:compton3D}). A photon incident on the camera undergoes Compton scattering in the first
detector, which records the location $x_1$ and the energy of interaction $E_1$. After the scattering, the photon is
absorbed in the second detector, where the position $x_2$ and energy of absorption $E_2$ are measured.
From the knowledge of $E_1$ and $E_2$, the scattering angle $\psi$ can be determined by the formula (e.g. \cite{ToddNightEverett,CreeBones})
\begin{equation}
 \cos \psi = 1-\frac{mc^2 E_1}{(E_1+E_2)E_2}.
\end{equation}
Here $m$ is the mass of an electron and $c$ is the speed of light. The direction into which a photon scatters is given by
 $\displaystyle-\mbeta:= \frac{\mx_2-\mx_1}{|\mx_2-\mx_1|}$. From the knowledge  of $\mbeta$ and the scattering angle
$\psi$ we conclude that the photon originated from the surface of the cone with central axis $\mbeta$, vertex $\mx_1$ and
opening angle $2\psi$. Therefore, although the exact incoming direction of the detected particle is not available, one knows
a cone of such possible directions. The goal of Compton camera imaging is to recover the distribution of the radiation sources from this data.

The description above addresses the Compton $\gamma$-cameras only. However, neutron detectors are currently being developed \cite{SpCharlton,shield} that (employing a different physics principle) provide the similar cone information. We thus will not make distinction between these and call them all ``Compton type cameras.''

We will now address briefly the mathematics behind the Compton type measurements.

\subsection{The Cone Transform}\label{S:CMP_math}
\begin{figure}[!ht]
\begin{center}
\scalebox{0.4}{\includegraphics{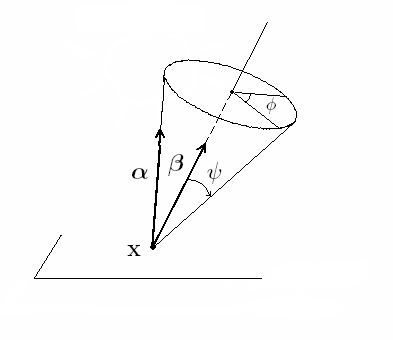}}
\caption{A Compton cone with apex $\mx$, central axis $\mbeta$ and half-angle $\psi$.}
\label{F:cone3D}
\end{center}
\end{figure}

%
Suppose that we want to image the distribution  $f(\my)$ of radioactivity sources inside a certain object in $\RR^3$.
From now on, we will assume that $f(\my)$ is supported on one side of the Compton camera and that $\mbox{supp} f$ does
not intersect the camera. Under the assumption of a sufficiently long observation time (and thus a large number of particles detected), the Compton camera provides us with the projections
\cite{CreeBones,BaskoZengGull98,TN_mathfound},
 which we denote by $\C f(\mx,\mbeta,\psi)$, i.e. the integrals of $f(\my)$ over cones parameterized by a vertex $\mx$ lying
 on the detector, central axis vector $\mbeta$ from the unit sphere $S^2$, and half-angle $\psi\in [0,\pi]$ (see Fig.\ref{F:cone3D}):
\begin{equation}\label{E:CMP_cone}
\C f(\mx,\mbeta,\psi) = K(\psi)  \int_0^{2\pi} \int_0^\infty f(\mx+r\malpha(\phi)) r \sin \psi \;dr\;d\phi,
\;\; \malpha\cdot\mbeta = \cos\psi,\;\;\malpha\in S^2.  
\end{equation}
Here $K(\psi)$ is the Klein-Nishina distribution of scattering angles\footnote{The Klein-Nishina factor is a known function and thus can be easily accounted for. We therefore do not include it in the further analysis.}. The function $\C f(\mx,\mbeta,\psi)$ provides the
 expectation of the number of hits, per unit time, from the particles moving along the cone. In particular, when the
particle count is low, these values are not well determined by the data. This issue was addressed in the preceding Sections.
An important observation is that the problem of inverting the cone transform  $\C f(\mx,\mbeta,\psi)$ is even more overdetermined than the $3D$ X-ray transform.
Indeed, $f(\my)$ is a function of three variables, while $\C f(\mx,\mbeta,\psi)$ depends on five parameters (here $\mx$ is
restricted to a 2D detector surface). Most authors consider restricted versions of the cone transform, reducing the number
of parameters from five to three \cite{CreeBones,BaskoZengGull98,TN_fixedaxis} . The paper \cite{Smith05} contains a closed form inversion formula, which uses four of the parameters\footnote{Recently \cite{MaxFranProst09}, the complete set of data was used for reconstructions.}.
We, however, have to use the full $5$-dimensional dataset, since, as it was mentioned before, any attempt to restrict the data would wipe out the signal.

\subsection{Reconstruction Techniques in Compton Camera Imaging}\label{SS:ComRec}

We provide here a brief survey of some known reconstruction methods in Compton cameras imaging. Many of them reduce cone projections to
Radon projections, i.e. integrals of $f$ over planes.

\emph{Algebraic Reconstruction Techniques}, as elsewhere in tomography,  are widely used in Compton cameras imaging. This is partially
 due to the fact that analytic inversion formulas were not available until the second half of 1990's. Maximum likelihood methods
 were developed in \cite{Singh_ML,Singh_iterative,Du_eval}. A maximum likelihood method was also used in imaging with the COMPTEL
 telescope \cite{Schonf_Comptel}. In \cite{Sauve} a matrix inversion technique was utilized to solve the problem for a specific
 detector geometry. Fast backprojection algorithms were developed in \cite{Rohe,Wilderman}.

A \emph{spherical harmonics expansion} solution was proposed in \cite{BaskoZengGull98}. For every point $\mx$ on the detector array, and a
fixed in advance half-angle $\psi$ the authors relate the conical projection $\C f(\mx,\mbeta,\psi)$ to the Radon projection
(integral) of $f$ over the plane passing through $x$ and perpendicular to $\beta$. A fast algorithm for computing the spherical
 harmonic series was also developed in this paper. Several other authors provided spherical harmonics solutions, however their model
for Compton data differs from (\ref{E:CMP_cone}), e.g. \cite{Parra,TomHir02,TomHir03}.

Probably the first \emph{closed form analytic inversion formula} was obtained in \cite{CreeBones}. The authors considered only cones perpendicular to
the detector plane, i.e. cones with central axis $\mbeta = \mz$. Thus, the Compton data depends only on the vertex $\mx:=(x,y)$ and
the opening angle $\psi$. Let us denote  $g(x,y,\psi):=\C~f(\mx,\mz,\psi)$ and let $F(u,v,\cdot),\; G(u,v,\cdot)$ be the two-dimensional
 $(x,y)$-Fourier transforms of $f(x,y,z)$ and $g(x,y,\psi)$ respectively. Then the following relation holds:
\begin{equation}
 F(u,v,\frac{\xi}{\sqrt{u^2+v^2}})=\mathcal{H}_0\left.\left[ \frac{u^2+v^2}{K(t)t\sqrt{1+t^2}}G(u,v,t)\right]\right|_{t=\xi},
\end{equation}
where $\xi = z\sqrt{u^2+v^2}$ and $t=\tan \psi$. Here $\mathcal{H}_0$ is the zero-order Hankel transform
\[\mathcal{H}_0 g(\rho)=2\pi\int_0^\infty g(r)rJ_0(2\pi r\rho)\;dr,\]
and $J_0(\gamma) = (2\pi)^{-1}\displaystyle\int_0^{2\pi} e^{i\gamma \cos\phi}\;d\phi$ is the zero-order Bessel function.
Later, in \cite{TN05radon}, more properties of the same restricted cone transform were given.

In \cite{MaxFranProst09} the authors use yet another mathematical formulation for the forward problem, which accounts
for the efficiency of the detector at different incident angles of particles. They extended the approach of  \cite{CreeBones} to show
that the radioactivity distribution can be reconstructed from any family of cones that have central axes $\beta$ such that the angle between $\beta$ and the $\mz$-axis is fixed. Then averaging of the solutions for different values of this angle was employed, thus making use of all available data and reducing noise.

The following useful relation between the cone and Radon transforms was found in \cite{Smith05}:
\begin{equation}
HRf(\mbeta, \mx\cdot\mbeta):=\frac{1}{\pi}\int_{-\infty}^\infty Rf(\mbeta,t)h(\mx\cdot\mbeta-t)\;dt=
-\frac{1}{\pi}\int_0^\pi \mathcal{C}f(\mx,\mbeta,\psi)h(\cos\psi)\;d\psi
\end{equation}
Here $H$ denotes the Hilbert transform and $R$ is the Radon transform in three dimensions. Note that
 four out of five variables are used and that the $HRf(\mbeta, \mx\cdot\mbeta)$ is constant on the line
 $L_c:=\{\mx\in\mbox{detector plane}|\mx\cdot\mbeta = c\}$, for any fixed constant $c$. The last observation shows that it is
possible to reconstruct $f$ if the vertices of cones $\mx$ are restricted to a curve. This also allows for averaging of the
solution on the lines $L_c$. In Section \ref{SS:CMP_3m} we will consider a two-dimensional analog of this formula.

\section{Compton Cameras in Two Dimensions}\label{S:2DCom}

\begin{figure}[!ht]
\begin{center}
\scalebox{0.5}{\includegraphics{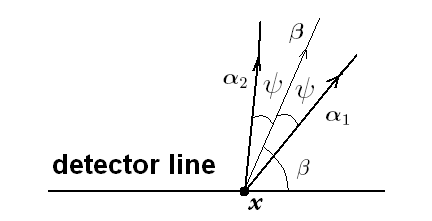}}
\caption{Two-dimensional cone with apex $\mx$, central axis $\mbeta$ and half-angle $\psi$.}
\label{F:compton2D}
\end{center}
\end{figure}

In the two dimensional setting we assume that the detectors $\mx$ lie on a line, which we call the {\em detector line}. A cone in two dimensions is defined by a point $\mx$ that serves as its vertex, a central axis unit vector $\mbeta$, and a half-angle $\psi$. Such cones simply consist of two rays with a common vertex. More precisely, let $\mbeta = (\cos\beta, \sin\beta) \in S^1, \beta \in [0,\pi]$ be the central axis of a cone with opening half-angle $\psi \in [0,\pi]$. The two half-lines that constitute the cone pass through the detector point $\mx$ and are given by
$$
\{\my\in \RR^2:\my =r\malpha_i, r\geq 0\}, i=1,2,
$$
where $\malpha_1=(\cos(\beta-\psi),\sin(\beta-\psi))$ and $\malpha_2=(\cos(\beta+\psi),\sin(\beta+\psi))$, see Fig. \ref{F:compton2D}. As we have agreed before, we drop the known factor $K(\psi)$. Thus, in what follows we will assume uniform distribution of scattering angles, that is $K(\psi)\equiv 1$.  Then the two dimensional cone transform $\C f(\mx,\mbeta,\psi)$ is the projection of $f(\my)$ onto these lines:
\begin{equation}\label{E:Compt2D}
 \C f(\mx,\mbeta,\psi) =  \int_0^\infty \left(f(\mx+r\malpha_1)+f(\mx+r\malpha_2)\right)\;dr = \D f(\mx,\malpha_1)+\D f(\mx,\malpha_2).
\end{equation}
Here $\D f(\mx,\alpha) = \int_0^\infty f(\mx+r\malpha)\;dr$ denotes the divergent beam (or fanbeam) transform of $f$.

For convenience, we will sometimes express $\C f$ and $\D f$ as functions of the angle $\beta$ instead of the unit vector $\mbeta$. Thus, we can write (\ref{E:Compt2D}) as
\begin{equation}\label{E:Compt2Da}
 \C f(\mx,\beta,\psi) = \D f(x,\beta-\psi)+\D f(x,\beta+\psi).
\end{equation}

As in $3D$, the problem of inverting the two-dimensional transform is overdetermined: $\C f(\mx,\mbeta,\psi)$ depends on three parameters, while $f(\my)$ only on two.

The authors are aware of few papers devoted to Compton cameras in two dimensions.  In \cite{BaskoZengGull97_1,BaskoZengGull_patent} subsets of the cone projections of $f(x)$ are represented as line integrals of the function
$f$ added with its mirrored shear transformation. Namely, let us introduce the
 variables $k_1 = \cot(\beta+\psi),k_2 = \cot(\beta-\psi)$. The cone transform can be written as
\begin{equation}
 \C f(x,\beta,\psi) = \int_0^\infty f(x+k_1 z,z)\;dz +\int_0^\infty f(x+k_2 z,z)\;dz,
\end{equation}
where the $x-$ axis coincides with the detector line and the $z-$ axis is perpendicular to the detectors. Let us fix the
sum $k_1+k_2=K$ and define the function
\begin{equation}
 f_K(x,z) =
\begin{cases}
  f(x,z),  & \mbox{if } z\geq 0 \\
  f(x-Kz,-z), & \mbox{if } z < 0.
\end{cases}
\end{equation}
Then, $\C f(x,\beta,\psi) = \int_{-\infty}^{\infty}f_K(x+k_1 z,z)$. Thus, for each fixed value of $K$, the cone
projections are represented as line integrals of a certain function, so all that remains is to invert the Radon transform.
The reconstructed image will contain the function $f$ above the detector line and a mirrored shear transformation of $f$
below the detectors. Obviously, by fixing the parameter $K$, not all available data is used. The case when $K=0$ corresponds
to using only those cones with central axes orthogonal to the detector line. Since we need to use all overdetermined data, this approach is inconvenient. Besides, as our previous consideration of tomographic methods shows, in our specific situation, backprojection methods, rather than attempting the ``full'' reconstruction should be used.

\subsection{Three Inversion Methods}\label{SS:CMP_3m}

\subsubsection{Method A}\label{SS:A}

The following relation, proven in \cite{Hristova}, makes use of all parameters and is an analog of the three dimensional result presented in \cite{Smith05}.

\begin{theorem} \label{T:inversion}
 Let the closed unit ball lie on one side of and away from the detector line, and let
$f(x)\in H_0^{1/2}(B^2)$. Denote $h(t)=1/t$. Then
\begin{equation}\label{E:Smith2D}
 H\R f(\mx\cdot \mbeta,\mbeta)= \frac{1}{\pi}\int_{-\infty}^\infty \R f(\mbeta,t)h(\mx\cdot\mbeta-t)\;dt=
-\frac{1}{\pi}\int_0^{\pi} \C f(\mx,\mbeta,\psi) h(\cos\psi)\;d\psi
\end{equation}
Here $H$ is the Hilbert transform in the first (scalar) variable and the integrals are understood in the principal value sense.
\end{theorem}

Here $B^2$ is the unit disk in the plane and $H^s_0$ is the standard notation for Sobolev spaces (e.g., \cite{Evans}).

\begin{corollary}\label{Cor:CMP}
 Theorem \ref{T:inversion} provides an inversion formula for the cone transform.
\end{corollary}
Indeed, computing the integral in the right hand side of (\ref{E:Smith2D}), one recovers $H\R f$, where $\R$ is the 2D Radon
transform, and $H$ is the Hilbert transform with respect to the linear variable. Then, the filtered backprojection formula for inversion
of the two-dimensional Radon transform,
\begin{equation}\label{E:FBP2d}
 f(x)= \frac{1}{4\pi}\R^\# \left(H\frac{\partial}{\partial s}\R f\right)(x),
\end{equation}
implies that after differentiating the right hand side of (\ref{E:Smith2D}) with respect to the linear variable
and backprojecting, one recovers the function $f$.

\begin{remark}
In the left hand side of (\ref{E:Smith2D}), one sees $\R f(s,\mbeta)$, where $s = \mx\cdot \mbeta$. Thus, if one infinite
detector array is used and $\mbeta$ is perpendicular to the array, we would only know $\R f(s,\mbeta)$ for a single value of $s$. If the
detector array is finite, as is the case in all implementations, we would miss a much bigger chunk of data. One possible way
to obtain the complete Radon data in $(s,\mbeta)\in[-1,1]\times S^1$ is, for example, to use three finite size detector arrays placed
along the sides of a square containing the object.
\end{remark}

\subsubsection{Method B}\label{SS:B}

Another, simpler way of obtaining Radon data from cone data is presented below. Let us denote by $u(\mx,\alpha)$ the integral
 of $f$ along the ray starting at the point $\mx$ in direction of the unit vector $\malpha$ (i.e. the fanbeam projection of $f$):
\begin{equation}\label{E:u}
u(\mx,\malpha)=\D f(\mx,\malpha):=\int_0^\infty f(\mx+r\malpha)\;dr.
 \end{equation}
Let us fix a detector position $\mx$ and a direction $\malpha_0$ and try to recover the fanbeam data $u(\mx,\malpha_0)$.
For any $\psi \in [-\pi,\pi]$ we can write $\C f(\mx,\beta,|\psi|) = u(\mx, \beta-|\psi|)+u(\mx, \beta+|\psi|)$. Thus, the
following relation holds
\[ u(\mx, \alpha_0) = \C f(\mx, \alpha_0 + \psi, |\psi|) -u(\mx,\alpha_0+2\psi) \]

\begin{figure}[!ht]
\begin{center}
\scalebox{0.6}{\includegraphics{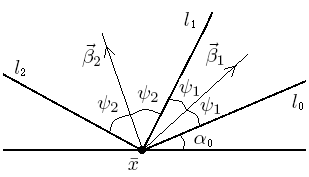}}
\caption{In order to determine the integral of $f$ along the line $l_0$, add the integral over the cone consisting of $l_0$ and
$l_1$ to the integral over the cone determined by $l_0$ and $l_2$. Then subtract the integral over the cone determined by $l_1$ and $l_2$.}
\label{F:methodB}
\end{center}
\end{figure}

It is easily seen now (Fig. \ref{F:methodB}) that for any two half-angles $\psi_1, \psi_2 \in [-\pi,\pi]$ the fanbeam data
$u(\mx,\alpha_0)$ can be obtained from cone data as follows:
\begin{equation}\label{E:Compt_av}
\begin{split}
 &u(\mx,\alpha_0) =\mbox{\hskip 350pt}\\
 & \mbox{\hskip 0.5in}\frac{1}{2}\left[\C f(\mx,\alpha_0+\psi_1, |\psi_1|)\right.+\\
 & \mbox{\hskip 1in}\left.\C f(\mx,\alpha_0+\psi_2, |\psi_2|)-\C f(\mx,\alpha_0+\psi_1+\psi_2, |\psi_1-\psi_2|) \right]
\end{split}
\end{equation}
The angles $\psi_1$ and $\psi_2$ were arbitrarily chosen, so one could average on $\psi_1$ and $\psi_2$ in order to use all
available data. This averaging also helps reducing the effects of the background noise.

If averaging over half-angles $\psi_1$ and $\psi_2$ is used, this method is computationally expensive. We propose
 a third reconstruction method, which is fast and makes use of all available data. \\

\subsubsection{Method C}\label{SS:C}

Let us recall that the source intensity function $f(\my)$ is compactly supported and $\mbox{supp} f$ lies on one side and away from the detector array. Let us fix a detector location $\ma$. Then the function $u(\ma,\alpha)$ (see (\ref{E:u})) is supported in $(0,\pi)$, so it can be
 extended periodically in $\alpha$ from $[0,2\pi]$ to $(-\infty,\infty)$.
 Let us define
\begin{equation}\label{E:u_tilde}
\begin{split}
 \tilde{u}(\ma, \alpha) := & \pi^{-1} \int_{-\alpha}^{\pi-\alpha} \C f(\ma,\alpha+\psi,|\psi|)\;d\psi
\end{split}
\end{equation}
We can write,
\begin{equation}
\begin{split}
 \tilde{u}(\ma, \alpha) =
\pi^{-1}\{\int_{-\alpha}^{\pi-\alpha}u(\ma,\alpha)\;d\psi+\int_{-\alpha}^{\pi-\alpha}u(\ma,\alpha+2\psi)\;d\psi\}=\\
u(\ma,\alpha)+(2\pi)^{-1}\int_{0}^{2\pi}u(\ma,\alpha)\;d\alpha
\end{split}
\end{equation}
In the above equation, the periodicity of $u(\ma,\alpha)$ with respect to $\alpha$ was used. Then,
\begin{equation}\label{E:approx}
\begin{array}{l}
 \tilde{u}(\ma,\alpha)  =
 u(\ma,\alpha)+\frac{1}{4\pi} \int_{|\momega|=1} \int_{-\infty}^\infty f(\ma + r \momega)\;dr d\momega\\ =
\D f(\ma,\alpha)+ (4\pi)^{-1}\left(\R^\# \R f\right)(\ma).
\end{array}
\end{equation}
Here, as before, $\R^\#$ denotes the backprojection operator
\[
\R^\#g(\my) = \int_{|\momega|=1}g(\my\cdot\momega,\momega)\;d\momega.
\]
The second term in the right-hand side of (\ref{E:approx}) can be written as
 \[
\D [(4\pi)^{-1}\R^\# \R f(\ma) \delta (\ma -\my)],
\]
where $\delta$ is the Dirac's delta-function.
Therefore, we come to the following conclusion:
\begin{theorem}\label{T:array_support}
When the inverse fan-beam transform is applied to $\tilde{u}$,
the result equals the sum of the function $f(\my)$ and a distribution supported on the detector array only. Hence, functions $f$ supported away from the detector arrays are recovered correctly.
\end{theorem}
In essence, this method approximates an integral of a function on a given line by averaging all its cone integrals which have one side lying on the line.

\section{$2D$ Compton data backprojection examples}\label{S:ComEx}
In order to generate Compton camera data, we first generated X-ray data in the manner described in Section \ref{SS:2DX}. Recall that the X-ray data for each particle consists of a bin on a detector array and exact incoming direction $\alpha$. Then, for every particle we chose a scattering angle $\psi$ drawn from a uniform distribution in $(-\alpha,\pi-\alpha)$.  The central axis of the scattering cone was computed as the sum $\beta = \alpha+\psi$. The information provided by the Compton camera is three dimensional and consists of the bin on the detector array, the central axis of the cone $\beta$ and the absolute value of the scattering angle $\psi$.

For the reconstructions from Compton camera data presented in this section, we used Method C, described in Section \ref{SS:C}, to convert the cone data into X-ray data. Method C was chosen as the least computationally expensive and also using all cone data. Note that for the particular type of data we are handling, the central axis and scattering angle of cones are continuous random variables. Thus, for a finite number detected particles, one cannot average on cones with a common side, because such cones are a rare occurrence. Therefore, the method is equivalent to simply treating a particle coming from a cone $(\mx,\beta,\psi)$ as two separate particles detected at $\mx$ and having incoming directions $\beta-\psi$ and $\beta + \psi$.
After converting cone data into X-ray data in this way, we employed the usual reconstruction procedures for X-ray data.

It is also important to make clear that in our experiments we assume that exact cone axes and scattering angles are known. The effects of uncertainty of these quantities are to be investigated in the future.

\begin{figure}[!ht]
\begin{center}
\scalebox{0.35}{\includegraphics{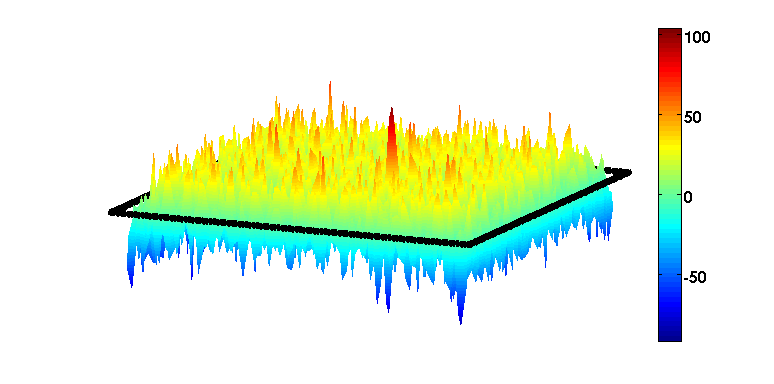}}
\scalebox{0.35}{\includegraphics{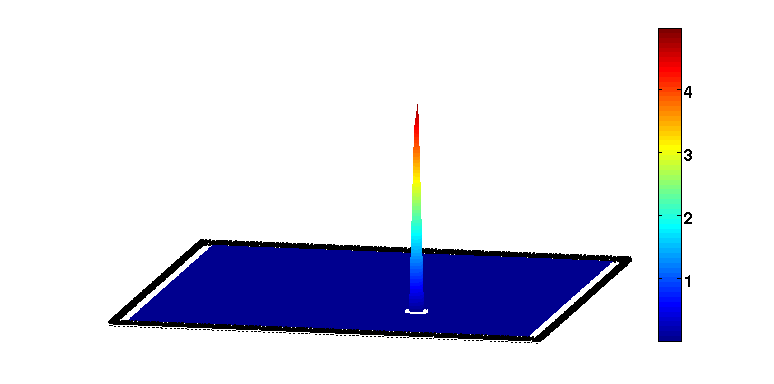}}
\caption{Filtered backprojection reconstruction from Compton data (left). The number of background particles is $900020$ versus $900$ source particles. The black lines represent
detector arrays. The right picture shows only the peaks which deviate more than $4.3$ standard deviations from the mean. The correct source location is found. Pixels close to the detectors are not shown, as the values of the reconstructed image there are quite high, see Theorem~\ref{T:array_support}.}
\label{F:conefbp_ex27}
\end{center}
\end{figure}

First, we used filtered backprojection to reconstruct the source distribution from data provided by four Compton detector arrays placed on the sides of the square $[-1,1]$. The signal to noise ratio was fixed at $0.1\%$ and we varied the bulk number of detected particles. As for the case of X-ray data, we defined a detection successful provided that highest peak of the reconstructed image occurred at the location of the source. The number of successful FBP detections increased slowly as the bulk number of detected particles went from $600,000$ to $1,000,000$. The results are reported in Table~\ref{T:cone}. An example of a successful detection using Method C combined with FBP is shown in Fig.~ \ref{F:conefbp_ex27}.

\begin{table}[!htb]
\label{T:cone}
\begin{center}
 \begin{tabular}{|c|c|c|c|c|c|}\hline
 method $\backslash$ bkgd particles & $600,000$ & $700,000$ & $800,000$ & $900,000$ & $1,000,000$\\\hline
  Method C + FBP               	 	 &  $8/20$   & $14/20$   &  $12/20$ & $18/20$ & $17/20$\\
  BP + local grids    		 &  $17/20$   & $17/20$   & $20/20$ & $20/20$ & $19/20$ \\\hline
\end{tabular}\\
\caption{Number of successful detections from cone data of a source located at $(0.311,-0.433)$, with a fixed SNR $0.1\%$ and varying number of detected particles, using different reconstruction methods. For each level of background particles $20$ sets of random data were generated. In each case we applied Method C combined with FBP and backprojection followed by local grids of size $9\times9$. In all but one experiment, if an FBP detection was successful, so was the corresponding backprojection detection. The one example in which FBP detection occurred but backprojection failed was at the level of $700,000$ background particles.}
\end{center}
\end{table}

\begin{figure}[!ht]
\begin{center}
\scalebox{0.35}{\includegraphics{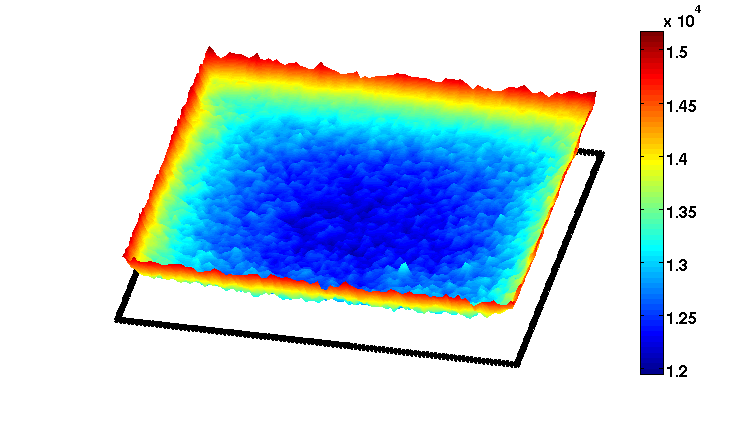}}\\
\scalebox{0.35}{\includegraphics{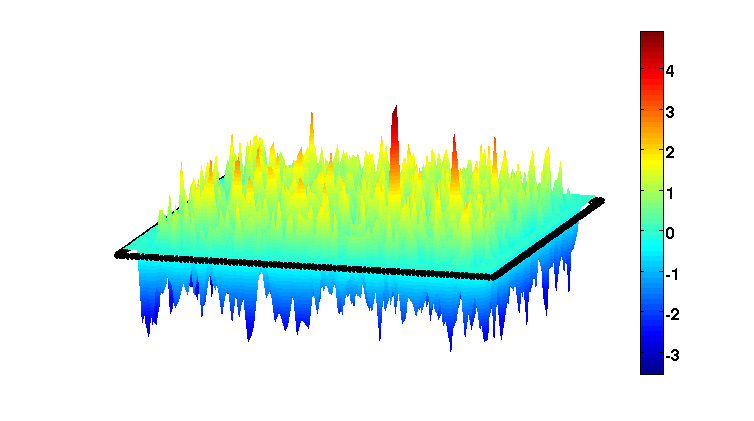}}
\scalebox{0.35}{\includegraphics{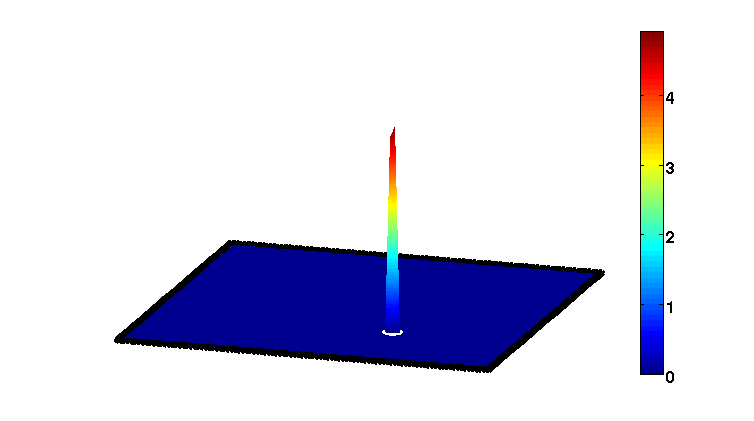}}
\caption{Unprocessed backprojection reconstruction from Compton data (top). The number of background particles is $900020$ versus $900$ source particles. The bottom left shows deviations from the (local) mean, measured in the units of (local) standard deviations. The size of local grids is $9\times 9$ pixels. Only the peaks that exceed the local mean more than $4.3$ (local) standard deviations are shown on bottom right. Source located at $(0.311,-0.433)$ is detected with confidence  $99.59\%$. }
\label{F:conebp_ex27}
\end{center}
\end{figure}

Next, we use Method C for converting cone data into X-ray data, followed by backprojection based on the line drawing algorithm described above. Thus, for every detected particle, two particles are backprojected along the sides of the cone. Therefore, the value of the resulting image at every pixel equals the number of cones intersecting it. On Fig.~\ref{F:conebp_ex27}(top) we show a typical backprojection reconstruction from four Compton detector arrays from exactly the same data used for the previous example (Fig.~\ref{F:conefbp_ex27}). The background is not uniformly distributed, so we employ the local grids approach introduced in Section~\ref{S:backproj}. The size of the local grids was chosen to be $9\times9$ pixels. On Fig.~\ref{F:conebp_ex27}(bottom) a plot of the number of local standard deviations away from the local mean is shown. The value at the source location deviates $4.93$ standard deviations from the mean, and the estimated confidence is $99.59\%$.

In Table~\ref{T:cone} we compared FBP reconstructions with backprojection reconstructions. For all backprojection results we used local grids of size $9\times9$ pixels on which the local means and standard deviations were estimated. In order to compare backprojection to FBP, we deemed a detection successful, if the number of local standard deviations was highest at the location of the source. Our results show that backprojection outperforms FBP, as in the case of X-ray data. The results are summarized in Table~\ref{T:cone}.

As we have already seen, detection based on backprojection allows for estimation of confidence. Similarly to detection by X-ray backprojection, one could set a threshold for detection at $4.1$ or $4.3$ local standard deviations above the local mean. According to our estimate for the confidence (\ref{E:CMP_conf}), such thresholds would result in confidences at least $81\%$ or $91.8\%$ respectively. In Table \ref{T:cone_conf} the number of successful detections for these two values of the threshold parameter are given. We used the same simulation data as for Table~\ref{T:cone}.

\begin{table}[!htb]
\label{T:cone_conf}
\begin{center}
 \begin{tabular}{|c|c|c|c|c|c|}\hline
 threshold $\backslash$ bkgd particles  & $600,000$ & $700,000$ & $800,000$ & $900,000$ & $1,000,000$\\\hline
  4.1 stds above mean          		&  $10/20$   & $9/20$   &  $14/20$ & $18/20$ & $18/20$\\
  4.3 stds above mean  		 	&  $7/20$   & $9/20$   & $11/20$ & $14/20$ & $16/20$ \\\hline
\end{tabular}\\
\caption{Number of successful detections from cone data with a fixed SNR $0.1\%$ and varying number of detected particles. Backprojection was used and  local means and standard deviations were estimated on local grids of size $9\times9$. Two different threshold values for detection were set, resulting in confidences at least $81\%$ and $91.8\%$ respectively.}
\end{center}
\end{table}

In comparison with X-ray data, a larger number of background particles is needed for detection with Compton cameras. This reflects the somewhat lower quality of Compton data in comparison with the data from collimated detectors.

A couple of examples of reconstructions from three Compton cameras, which form a "gate" of detectors, are provided below.

\begin{figure}[!ht]
\begin{center}
\scalebox{0.35}{\includegraphics{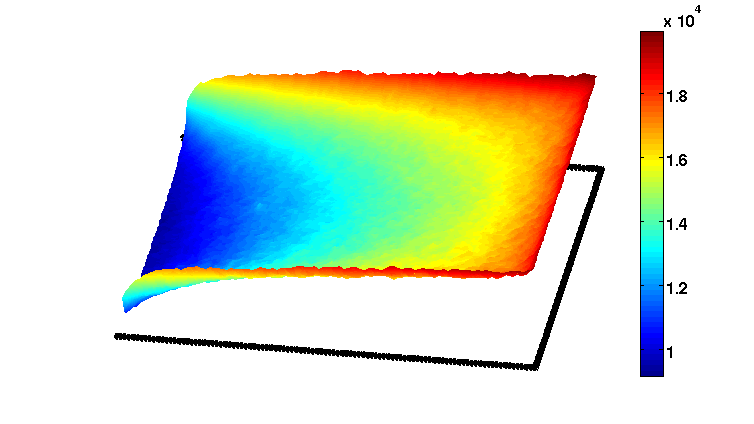}}\vspace{1cm}\\
\scalebox{0.35}{\includegraphics{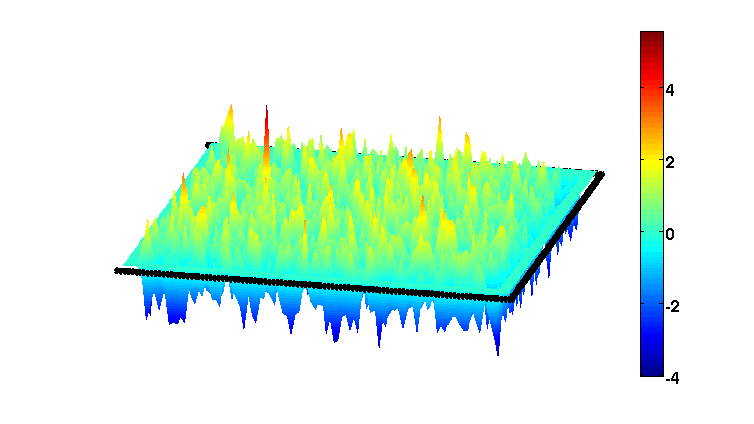}}
\scalebox{0.35}{\includegraphics{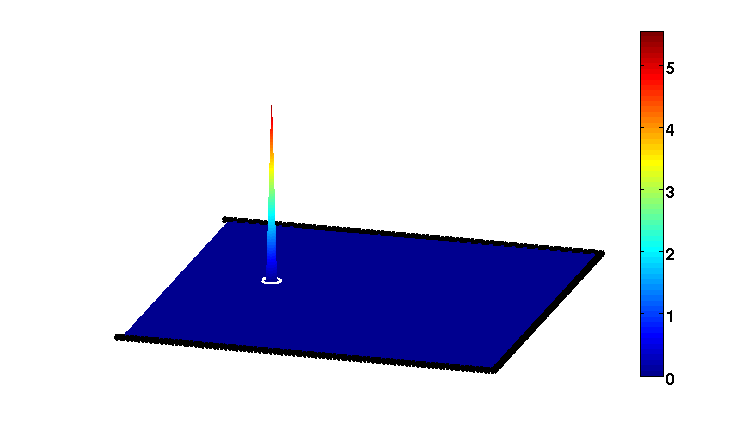}}
\caption{Backprojection reconstruction from Compton data (top). Detected source particles - $1050$, background particles - $999,925$, source location - $(-0.503,0.11)$.  Number of (local) standard deviations above the (local) mean at each pixel is shown on bottom left. The size of local grids is  Peaks that deviate more than $4.3$ (local) standard deviations from the mean are shown on bottom right. The source is detected with confidence $99.98\%$.}
\label{F:CMP_ComBP3}
\end{center}
\end{figure}

The source in Fig.~\ref{F:CMP_ComBP3} is close to the side of the square where there is no detector, which makes it harder to observe. The number of ballistic source particles was $1050$ versus $999,925$ background particles. The local mean and standard deviation were estimated on $9\times9$ local grids. The maximum number of local standard deviations was $5.56$ and occurred at the location of the source. The estimated confidence of detection is $99.98\%$.

\begin{figure}[!ht]
\begin{center}
\scalebox{0.35}{\includegraphics{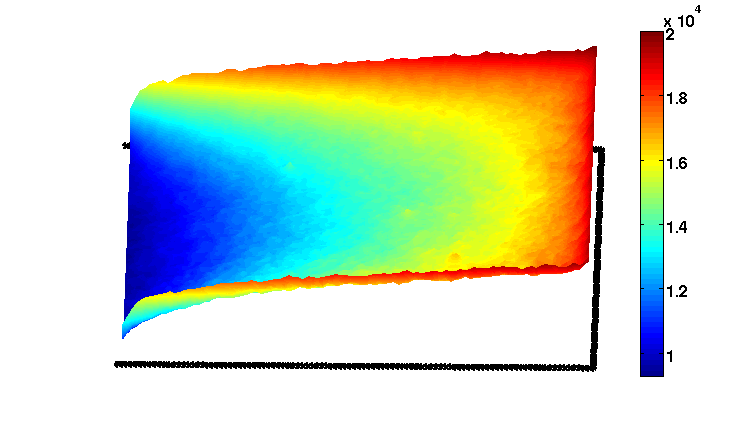}}\vspace{1cm}\\
\scalebox{0.35}{\includegraphics{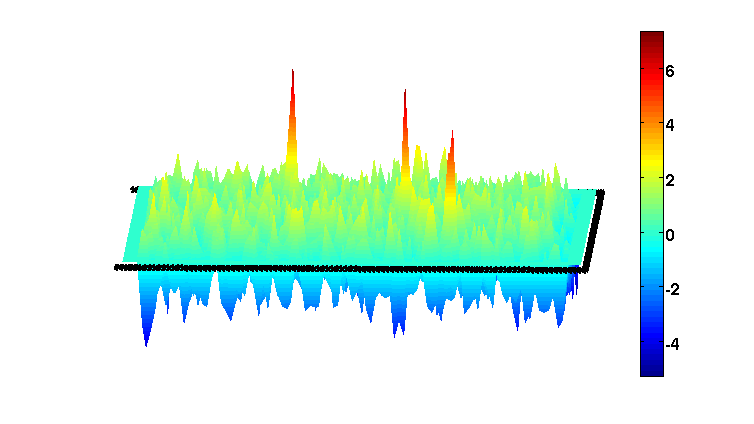}}
\scalebox{0.35}{\includegraphics{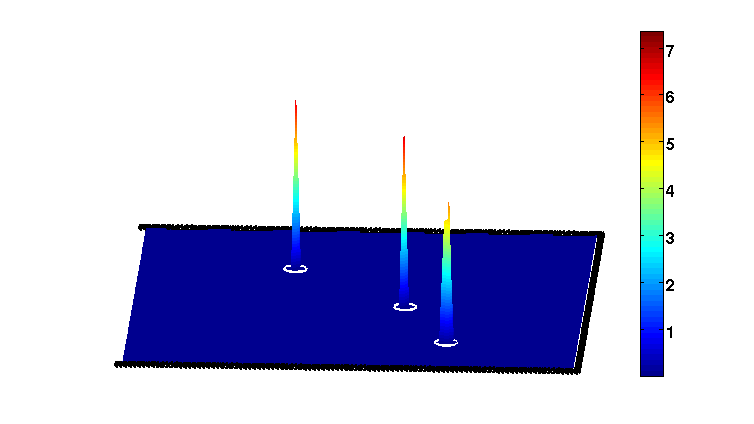}}
\caption{Backprojection reconstruction from Compton data (top). Sources located at $(0.203,-0.101)$, $(-0.301,0.43)$ and $(0.41,-0.61)$ emitted $1108$, $981$ and $1140$ particles, correspondingly. The number of detected background particles was $999,325$. Number of (local) standard deviations above the (local) mean at each pixel is shown on bottom left. Peaks that deviate more than $4.3$ (local) standard deviations from the mean is shown on bottom right. Confidence of detection based on the threshold $k_t = 4.3$ is $91.97\%$.}
\label{F:CMP_3sources}
\end{center}
\end{figure}

Finally, we present an example of a successful detection of several sources (Fig.~\ref{F:CMP_3sources}). Three sources were placed inside the imaged area, and three Compton-type detector arrays were used to detect particles.  The total number of detected ballistic source particles was $3,229$, each of the sources having roughly the same contribution. Additional $999,325$
random background particles were detected. We have set up the threshold to $k_t = 4.3\sigma$, which gives confidence of detection at least $91.97\%$. Note that this confidence probability is estimated based on the threshold, and not on the actual number of deviations at each peak, as in the previous examples. Thus, individual confidence probabilities for the three sources discovered are even higher. The local grids we used for estimating the local means and standard deviations had dimensions $9\times 9$.

\section{Final remarks and conclusions}\label{S:remarks}

Here are the main conclusions that we have reached:
\begin{enumerate}
  \item Direction sensitive detectors are needed to be able to uncover existence of a low emission source in the presence of dominating background noise. Collimated cameras, although providing the most valuable direction information, are not applicable, since collimation would decimate the already low signal. Compton type cameras can be used instead.
  \item Although standard integral-geometric models of emission tomography (X-ray and Radon transforms and their attenuated versions) are not applicable to the situation of a very weak source, they work to some extent in the detection problem. However, they fail way before reaching the low values of SNR needed in applications.
  \item In fact, only the backprojection part of the tomographic reconstruction does the detection work, and the filtering part worsens the results.
  \item The (unknown) attenuation might lower the SNR, but otherwise is irrelevant for the validity of the backprojection method.
  \item The conclusions above hold both for the collimated and Compton type cameras.
  \item The backprojection procedure can be justified by a simple probabilistic consideration, which also provides confidence probabilities of detection.
  \item A simple algorithm is provided that recovers the source from Compton data at SNRs on the order $0.1\%$ and beyond. It also provides confidence probabilities. In theoretical considerations and numerical tests, the algorithm demonstrates high sensitivity and specificity. The procedure can be made completely automatic, without the need of a human eye assessment of the image.
  \item The algorithm is based upon the assumption of an uniform random background. However, the non-uniformity that arises when the object is only partially surrounded by detector arrays, can be handled successfully by using local grids.
\item
As it is done in SPECT, one can try to use Bayesian methods to approach the same detection problem. This was done in $2D$ case in \cite{Xun}.
\end{enumerate}

Certainly, there remain quite a few questions that need to be resolved. Some of them will be discussed in the next publication. In particular, $3D$ Compton backprojection algorithms will be developed and studied analytically and numerically; more complex cargo structures will be modeled and tested, which will involve non-uniform attenuation and scattering; local grid version of the algorithm will be tested on more complex non-uniform random backgrounds than the ones arising due to the partial view; dependence of the results on the precision of Compton type cameras will also be addressed. We assumed that the exact incoming direction of particles is detected. In practice this is usually not the case, as only a probability distribution of the directions may be known. The effect of directional uncertainty is important from practical standpoint and we plan to consider it in the future.

\section*{Acknowledgments}
Work of D.~D., G.~K., and P.~K. was partly supported by the DHS Grant 2008-DN-077-ARI018-04. Work of M.~A., D.~D., Y.~H., and P.~K. was partly supported by the NSF DMS grants 0604778 and 0908208. Work of M.~A., G.~K., and P.~K. was supported in part by the IAMCS (Institute for Applied Mathematics and Computational Science of Texas A\&M University).  Y.~H. thanks the Institute for Mathematics and Its Applications (IMA) for the support and hospitality. Research at the IMA is supported by the National Science Foundation and the University of Minnesota. Y.~H. and P.~K. also thank the MSRI, where part of this work was done.

Thanks also go to our colleagues W.~Bangerth, R.~Carroll, W. Charlton, Sunil S. Chirayath, D.~Cline, B.~Mallick, G.~Spence, J.~Ragusa, S.~Subba~Rao, X.~Xun, and J.~Zinn for useful discussions and information.



\end{document}